\documentclass[ aps
               ,prl
               ,floatfix
	           ,footinbib
	           ,a4paper
	           ,superscriptaddress
	           ,amsmath
	           ,amssymb
               ,preprint,endfloats
               ]{revtex4}

\usepackage{rotating}
\usepackage{amsmath}
\usepackage{graphics}
\usepackage{textcomp}
\usepackage{dcolumn}
\usepackage{bm}
\usepackage{gensymb}
\usepackage{makecell}
\usepackage[plain,cm]{fullpage}

\begin{document}

\title{Ab-initio elastic tensor of cubic Ti$_{0.5}$Al$_{0.5}$N alloy:
the dependence of the elastic constants on the size and shape of the supercell model}

\author{Ferenc Tasn\'adi}
\email{tasnadi@ifm.liu.se}
\author{M. Od\'en}
\author{Igor A. Abrikosov}
\affiliation{Department of Physics, Chemistry and Biology
(IFM), Link\"oping University, SE-581 83 Link\"oping, Sweden}

\date{\today}
\begin{abstract}
In this study we discuss the performance of approximate SQS supercell models in describing the cubic
elastic properties of B1 (rocksalt) Ti$_{0.5}$Al$_{0.5}$N alloy by using a symmetry based projection
technique. We show on the example of  Ti$_{0.5}$Al$_{0.5}$N alloy, that this projection technique
can be used to align the differently shaped and sized SQS structures for a comparison in modeling
elasticity. Moreover, we focus to accurately determine the cubic elastic constants and
Zener's type elastic anisotropy of Ti$_{0.5}$Al$_{0.5}$N. Our best supercell model, that captures
accurately both the randomness and cubic elastic symmetry, results in $C_{11}=447$ GPa, $C_{12}=158$
GPa and $C_{44}=203$ GPa with 3\% of error and $A=1.40$ for Zener's
elastic anisotropy with 6\% of error. In addition, we establish the general importance of selecting
proper approximate SQS supercells with symmetry arguments to reliably model elasticity of alloys.
In general, we suggest the calculation of nine elastic tensor elements
- $C_{11}$, $C_{22}$, $C_{33}$, $C_{12}$, $C_{13}$, $C_{23}$, $C_{44}$, $C_{55}$ and $C_{66}$,
to evaluate and analyze the performance of SQS supercells in predicting elasticity of cubic alloys
via projecting out the closest cubic approximate of the elastic tensor. The here described
methodology is general enough to be applied in discussing elasticity of substitutional alloys with
any symmetry and at arbitrary composition.
\end{abstract}
%
\maketitle
%
%
\section{Introduction}
TiAlN coatings with their good oxidation resistance and excellent mechanical properties have
attracted high technological and academic interest \cite{Horling2005}. Several studies have been
devoted to discuss these alloys from different aspects to extend our understanding in maximizing
their functionality and operational efficiency. The thermodynamics, phase stability and spinodal
decomposition in TiAlN have been analyzed \cite{Mayrhofer_APL2006,Alling2007}, also on the influence
of nitrogen off-stoichiometry \cite{Alling2008} and pressure \cite{Alling2009}. Furthermore, the
theoretical prediction of the mixing enthalpy when alloying TiAlN with Cr has resulted in a general
design route to improve the thermal stability of hard coatings \cite{Lind2011}. Recently, the
importance of the significant elastic anisotropy in TiAlN on the isostructural spinodal decomposition
has been disccussed \cite{Tasnadi2010,Abrikosov2011}.
Though, the available theoretical tools with the help of modern supercomputers allows us to
tackle such complex physical phenomena in alloys \cite{Ruban2008}, the prediction of anisotropic
tensorial materials properties of substitutional alloys from first principles remains a
challenging and highly requested task in computational materials science \cite{Liu2005,VandeWalle2008}.
For example, in dynamical simulations suitably designed simulation
cells can greatly reduce the computational costs of predicting the temperature dependence of the
elastic and piezoelectric tensors of alloys.
The importance of the elastic and piezoelectric tensors of materials can be underlined not only by
its fundamental role in materials science but also their distinguished usage in (micro)mechanical
modeling, engineering or designing of machine elements, sensors, telecommunication devices, aircrafts,
etc..

Although the ordinary scalar cluster expansion \cite{Asta1991} offers an exact treatment of the
thermodynamics of alloys and its tensorial generalization \cite{VandeWalle2008} gives the most
elegant description of anisotropic tensorial materials quantities of alloys, the computationally
less demanding and less complex special quasirandom structure (SQS) approach \cite{Zunger1990}
is {\it more favorized} due to its simplicity and success. For example the giant piezoelectric
response of ScAlN alloys \cite{Tasnadi2010PRL,Wingqvist2010} or the mechanical properties of
TiAlN \cite{Tasnadi2010} have been successfully described within this approach. Using different
superstructures, Mayrhofer et al. have discussed the impact of the microscopic configurational
freedom on the structural, elastic properties and phase stability in TiAlN \cite{Mayrhofer2006}.
In B-doped wurtzite AlN significant configurational dependence of the piezoelectric constant has
been predicted \cite{Tasnadi2009} with presuming wurtzite symmetry, similarly to the discussion
of electronic properties and nonlinear macroscopic polarization in III-V nitride alloys
\cite{Mader1995,Bernardini2001}.

In fact, in these studies the success of the SQS approach in describing the energetics of alloys
is presumed for predicting tensorial materials properties when using different approximate
SQS supercells or even ordered structures. These works were either only predictive on the materials
constants or the confirmation of the applied approximate structural model was based on the
experimental agreement of the results. Moreover, most of the previous theoretical works on predicting
elasticity and piezoelectricity of alloys presumed the experimentally observed symmetry for the
modeling SQS supercells, though the substitutional disorder of the atoms in general breaks the
local point symmetry of the supercell, and focused only on the corresponding principal symmetry
non-equivalent tensor elements. While the symmetry arguments based tensorial version of the
cluster expansion \cite{VandeWalle2008} gives an exact approach to completely include the local
point symmetry of the materials, improperly chosen SQS supercells may result in large discrepancy
between theory and experiments or in erroneous theoretical findings. 

The SQS approach, in principle, is not aimed to generate structures with the inclusion of local
point symmetry and thus to provide the proper, full description of tensorial properties of alloys.
In fact, different SQS supercells break the symmetry somewhat differently and thus the comparison
of the differently shaped and sized SQS supercells in terms of modeling the elasticity of cubic
Ti$_{0.5}$Al$_{0.5}$N is a rather complex issue. Hence, detailed systematic studies on the
application of SQS supercells in predicting elastic constants of alloys are required to establish
their performance and to determine their applicability limits. For example, J. von Pezold {\it et}
al. \cite{VonPezold2010} have recently evaluated the performance of symmetricaly shaped
- ($A \times A\times A$), supercells for the description of elasticity in substitutional AlTi
alloys and obtained convergence and error bars for the {\it cubic-averaged} principle cubic elastic
constants within the supercell configuration space. However, a general concept of comparing and
measuring different sized and shaped SQS supercells in describing tensorial materials properties
is still lacking.

In this study, we present a general projection approach to establish a way of comparing the
{\it ab-initio} calculated elastic constants of B1 Ti$_{0.5}$Al$_{0.5}$N obtained with applying
different sized and shaped SQS supercells. We accurately predict and extensively discuss the
calculation of the principal cubic elastic constants of B1 Ti$_{0.5}$Al$_{0.5}$N within the
SQS approach. In general, we establish the importance of selecting proper SQS supercells with
symmetry arguments to reliably model elasticity of alloys. Namely, we show that supercells
even with good short range order (SRO) parameters may result in large non-cubic elastic constants
and, on the contrary, supercells with bad SRO parameters might approximate cubic elastic symmetry
fairly accurately. We give the convergence of elasticity with respect to different SQS supercells
for B1 Ti$_{0.5}$Al$_{0.5}$N via the symmetry projected cubic elastic constants. Moreover, we
suggest the calculation of 9 elastic tensor elements - $C_{11}$, $C_{22}$, $C_{33}$, $C_{12}$, $C_{13}$,
$C_{23}$, $C_{44}$, $C_{55}$ and $C_{66}$, instead of 21, to evaluate and analyze the performance
of SQS supercells in predicting elasticity of cubic alloys.
%
%
\section{Method}
In this section we provide a description of the techniques we applied to calculate and analyze the
approximate elastic constants of cubic B1 Ti$_{0.5}$Al$_{0.5}$N alloy. First we explain the applied
special quasirandom structure (SQS) approach of modeling alloys and discuss the difficulties of
describing the proper symmetric tensorial materials constants within the model. After that, we
summarize the computational details of obtaining the energetics and extracting the elastic tensors
of the approximate SQS supercells. Finally, we present a general projection method that provides a
technique to compare and analyze the calculated approximate elastic tensors and what establishes a
principle to discuss the supercell models in terms of modeling elasticity with the inclusion of
local point symmetry. The here described methodology is general enough to be applied for substitutional
alloys with any symmetry and arbitrary composition.
\subsection{Special quasirandom structure approach and its symmetry}
The special quasirandom structure (SQS) approach \cite{Zunger1990} greatly reduces the computational
difficulties of modeling thermodynamics, mechanical and electronic materials properties of random
alloys. The approach models the substitutional disordered alloys with ordered superstructures.
The basic structural element of the SQS model is a supercell, what is aimed to capture the structural
short range order (SRO) in alloys while its periodic repetition introduces spatial long range order
(LRO) \cite{Mader1995}. The degree of SRO is usually measured by the Warren-Cowley parameter
\cite{Cowley1950}, which for a pseudobinary A$_{1-x}$B$_x$N alloy is defined as $\alpha_j=1-P_B(R)/x_B$,
where $P_B(R)$ is the probability of finding a $B$ atom  at a distance $R$ from an $A$ atom and $x_B$
stands for the concentration of $B$. A perfectly random alloy is characterized by vanishing SRO,
while $\alpha>0$ and $\alpha<0$ define clustering and ordering, respectively. In terms of modeling
disorder, approximate SQS supercells with with small or vanishing SROs up to a certain neighboring
order can be compared if and only if the interaction parameters are also known. In this work the
atomic configurations in the supercells were obtained by including the Warren-Cowley SRO parameters
of the first seven nearest-neighboring shells. Namely, the disorder has been considered up to the
seventh neighboring shell on the metal sublattice. Accordingly, the SRO parameters were calculated
only on the Ti-sublattice. In order to achieve the closest possible model of the perfectly random
alloy in the chosen sized and shaped supercell approximation $(A\times B\times C)$, a Metropolis-type
simulated annealing algorithm \cite{Metropolis1953} has been applied with a cost-function built
from the properly weighted nearest-neighbor SRO parameters.

The SQS supercell approach in general breaks the local point symmetry at different stages.
The substitutional disorder changes the microscopic local environments which results also in some
distorsions on the lattice parameters. Namely, after full relaxation the supercells will have a
general triclinic shape. Moreover, the SQS approach in modeling the substitutional disorder of
alloys allows one to apply arbitrary supercell shape and size - $(A\times B\times C)$ in terms of
lattice vectors. This arbitrariness though increases the variational freedom to obtain closely
vanishing SRO parameters with relative small supercell size at any alloy composition, it also
spoils the symmetry of the model. Thus, the elasticity of the B1 Ti$_{0.5}$Al$_{0.5}$N alloy is modeled
with fully relaxed SQS supercells an it is described by 21 elastic constants, instead of the three
principal cubic constants, $C_{11}$, $C_{12}$ and $C_{44}$. Namely, in the SQS approach the
elastic tensor of the model belongs to a symmetry class that is lower than the one that the alloy
shows experimentally. Furthermore, different SQS supercells break the symmetry somewhat differently,
which means that the comparison of the results can only be done after certain alignment. In this
study we show that a projection technique can provide such an alignment in the example of the B1
Ti$_{0.5}$Al$_{0.5}$N alloy.
\subsection{Calculational technique to obtain the elastic tensors}
To obtain total energies and extract the elastic constants of the supercells introduced above,
Density Functional Theory (DFT) calculations have been performed with using the plane-wave
ultrasoft pseudo-potential \cite{Vanderbilt1990} based Quantum Espresso program package
\cite{Giannozzi2009}. The exchange correlation energy was approximated by the Perdew-Burke-Ernzerhof
generalized gradient functional (PBE-GGA) \cite{Perdew1996}. The plane-wave cutoff energy together
with the Monkhorst-Pack sampling \cite{Monkhorst1976} of the Brillouine zone were tested and
sufficient convergence was achieved. The pseudopotentials were downloaded from the library linked
to Quantum Espresso and tested by calculating the elasticity of bulk B1 AlN and TiN in agreement
with literature values \cite{Chen2003,Mayrhofer2006}. In obtaining the ground state structure of
the modeling supercells, both the lattice parameters and the internal atomic coordinates were
relaxed by using the extended molecular dynamics method with variable cell shape introduced
by Wentzcovitch \cite{Wentzcovitch1993}. Accordingly, during the relaxation the supercells
geometries have been changed from the initial cubic-like lattice structure and converged to a
slightly distorted triclinic shape with vanishing stress tensor. Thus, we avoid any residual
structural stresses, which is essential in performing an accurate comparative analysis of the
calculated elastic tensors. In this dynamics, a value of 0.02 KBar was taken as convergence threshold
for the pressure. The elastic constants were calculated via the second order Taylor expansion
coefficients of the total energy
\begin{equation}
C_{ij}=\frac{1}{V_0}\frac{\partial^2E(\epsilon_1,\ldots,\epsilon_6))}{\partial\epsilon_i\partial\epsilon_j}\Big|_0
\end{equation}
where Voigt's notation is used to describe the strain $\epsilon$ and elastic $C_{ij}$ tensor
\cite{Vitos2010,Nye1985}.
To obtain the entire elastic tensor, namely the 21 elastic constants of each supercell, 21
different distorsions have been applied without volume conservation. The elastic constants were
calculated by standard finite difference technique from total energy data obtained from
$\pm$ 1\% and $\pm$2\% distorsions.
\subsection{Projection of the elastic tensor to the closest elastic tensor of higher symmetry}
In this section we describe the projection technique introduced by Moakher et al. \cite{Moakher2006}
to obtain the closest elastic tensor with higher symmetry class for any given elastic tensor with
arbitrary symmetry. This projection technique allows us to extract the largest cubic part of the
calculated elastic tensors. It introduces a tool to compare the obtained approximate elastic tensors
and measure the appropriateness of the SQS supercells in modeling the elasticity of B1
Ti$_{0.5}$Al$_{0.5}$N. 

The symmetric elastic tensor has 21 inequivalent elements for the most general triclinic system.
A system with higher point symmetry requires less parameters in describing its elastic behavior. For
example, with cubic symmetry the material has only 3 principal elastic constants,
$C_{11}, C_{12}$ and $C_{44}$, while the hexagonal point symmetry results in 5 elastic constants,
$C_{11}, C_{12}, C_{13}, C_{33}$ and $C_{44}$. Nevertheless, any elastic tensor can be expressed
as a vector in a 21 dimensional vector space, with the following components
\begin{eqnarray}
&E=(C_{11},C_{22},C_{33},\sqrt{2}C_{23},\sqrt{2}C_{13},\sqrt{2}C_{12},2C_{44},2C_{55},\nonumber\\
   &2C_{66},2C_{14},2C_{25},2C_{36},2C_{34},2C_{15},2C_{26},2C_{24},\nonumber\\
   &2C_{35},2C_{16},2\sqrt{2}C_{56},2\sqrt{2}C_{46},2\sqrt{2}C_{45}),
\label{elastic_vector}
\end{eqnarray}
where the $\sqrt{2}$'s ensure the invariance of the norm on the representation, whether it
is vector or matrix. For the basis vectors see Ref.\cite{Browaeys2004}.
The following projectors $P_{\text{sym}}$ generate the closest
elastic tensor with higher symmetry via
\begin{equation}
E^{\text{sym}}=P^{\text{sym}}E,
\end{equation}
where $E_{\text{sym}}$ has higher point symmetry.
The term closest here is used in the sense, that the Euclidean distance $||E-E^{\text{sym}}||$ is minimum.

To obtain the closest cubic approximate in our study of B1 Ti$_{0.5}$Al$_{0.5}$N, we applied the projector
given as a 21$\times$21 matrix,
\begin{eqnarray}
&&P^{\text{cub}}=\begin{pmatrix}p_{\text{cub}}&0_{9\times 12}\\0_{12\times 9}&0_{12\times 12}\end{pmatrix},
\nonumber\\[3mm]
&&p^{\text{cub}}=\begin{pmatrix}
1/3&1/3&1/3&0&0&0&0&0&0\\
1/3&1/3&1/3&0&0&0&0&0&0\\
1/3&1/3&1/3&0&0&0&0&0&0\\
0&0&0&1/3&1/3&1/3&0&0&0\\
0&0&0&1/3&1/3&1/3&0&0&0\\
0&0&0&1/3&1/3&1/3&0&0&0\\
0&0&0&0&0&0&1/3&1/3&1/3\\
0&0&0&0&0&0&1/3&1/3&1/3\\
0&0&0&0&0&0&1/3&1/3&1/3\\
\end{pmatrix}.\nonumber\\
\label{cubic_proj}
\end{eqnarray}
Accordingly, the projected cubic elastic constants can be achieved via the following
simple averaging,
\begin{eqnarray}
\bar{C}_{11}&=&\frac{C_{11}+C_{22}+C_{33}}{3}\nonumber\\
\bar{C}_{12}&=&\frac{C_{12}+C_{13}+C_{23}}{3}\nonumber\\
\bar{C}_{44}&=&\frac{C_{44}+C_{55}+C_{66}}{3}.
\label{cubic_average}
\end{eqnarray}
We can call them cubic-averaged elastic constants, since the equation is equivalent with averaging
over the three orthogonal directions, [100], [010] and [001]. We note here that this averaging was
used by von Pezold {\it et} al. in searching for optimized supercell in AlTi alloys.
Thus, to obtain the closest cubic projection of an elastic tensor with arbitrary symmetry one
needs to derive 9 different distorsion and calculate 9 independent tensor elements, like
$C_{11}, C_{22}, C_{33}, C_{23}, C_{13}, C_{12}, C_{44}, C_{55}$ and $C_{66}$. In case of
cubic symmetry Eq.(\ref{cubic_average}) results in the well-known cubic identities of the
elastic constants, see Eq.(\ref{cubic_sym}).
\begin{widetext}
For modeling elasticity in hexagonal alloys, one needs the closest hexagonal approximation that can
be obtained via the projector
\begin{eqnarray}
P^{\text{hex}}=\begin{pmatrix}p_{\text{hex}}&0_{9\times 12}\\0_{12\times 9}&0_{12\times 12}\end{pmatrix},\quad
p^{\text{hex}}=\begin{pmatrix}
3/8&3/8&0&0&0&1/(4\sqrt{2})&0&0&1/4\\
3/8&3/8&0&0&0&1/(4\sqrt{2})&0&0&1/4\\
0&0&1&0&0&0&0&0&0\\
0&0&0&1/2&1/2&0&0&0&0\\
0&0&0&1/2&1/2&0&0&0&0\\
1/(4\sqrt{2})&1/(4\sqrt{2})&0&0&0&3/4&0&0&-1/(2\sqrt{2})\\
0&0&0&0&0&0&\frac{1}{3}&\frac{1}{3}&0\\
0&0&0&0&0&0&\frac{1}{3}&\frac{1}{3}&0\\
1/4&1/4&0&0&0&-1/(2\sqrt{2})&0&0&1/2
\end{pmatrix},\nonumber\\
\end{eqnarray}
that acts in the same 9 dimensional subspace and
results in the following expressions for the projected hexagonal elastic constants,
\begin{eqnarray}
&&\bar{C}_{11}=3(C_{11}+C_{22})/8+C_{12}/4+C_{66}/2,\quad
\bar{C}_{12}=(C_{11}+C_{22})/2+3C_{12}/4-C_{66}/2,\nonumber\\
&&\bar{C}_{13}=(C_{13}+C_{23})/2,\quad
\bar{C}_{33}=C_{33},\quad
\bar{C}_{44}=(C_{44}+C_{55})/2.
\end{eqnarray}
\end{widetext}
A detailed derivation of the projectors for the all symmetry classes, monoclinic,
orthorombic, tetragonal, trigonal, hexagonal, cubic and isotropic can be found in Ref. \cite{Browaeys2004,Moakher2006}.
It is worth to mention that not all projectors can be defined in the above used 9
dimensional subspace. Furthermore, the application of this projection technique allows one
to spilt the elastic tensor into a direct sum of tensors with different symmetry.
Such decomposition is possible, for example, on the following routes,
\begin{eqnarray}
E&=&E_{\text{cubic}}+E_{\text{tetragonal}}+E_{\text{orthorombic}}\nonumber\\&&+E_{\text{monoclinic}}+E_{\text{triclinic}}\nonumber\\
E&=&E_{\text{hexagonal}}+E_{\text{tetragonal}}+E_{\text{orthorombic}}\nonumber\\&&+E_{\text{monoclinic}}+E_{\text{triclinic}}.
\end{eqnarray}
Thus, with calculating the norm of the components one gets information about the different contributions
and can analyze elastic anisotropy in general \cite{Browaeys2004}.
%
\section{Results and discussion}
In this section we present a comparative analysis of the calculated approximate elastic tensors
obtained for the cubic (B1) TiA$_{0.5}$l$_{0.5}$N alloy within the special quasirandom structure
approach. To get different levels of the approximation of the elasticity in cubic Ti$_{0.5}$Al$_{0.5}$N,
several approximate SQS supercell models have been generated with different shape and size, such as
$(2\times2\times2)$, $(2\times3\times2)$, $(4\times3\times2)$, $(4\times3\times4)$, $(4\times4\times3)$
and $(4\times4\times4)$. Here SQS supercell sizes are measured in terms of the fcc unit vectors.
To have a more complete comparison of the calculated elastic tensors, we present results obtained with
the ordered L1$_0$ structure and three other structures, denoted here by C1-$(2\times2\times2)$,
C3-$(2\times2\times2)$ and B1-$(2\times2\times2)$. These three structures are not based on the fcc
unit cell but on the fcc Bravais cell. The C1-$(2\times2\times2)$ and C3-$(2\times2\times2)$ was
created by Mayrhofer et al. \cite{Mayrhofer2006} with considering the number of bonds between the
host and doping atoms. The C3-$(2\times2\times2)$ structure was designed with preserving the cubic
symmetry. The B1-$(2\times2\times2)$ structure was obtained by von Pezold with using a Monte-Carlo
scheme and averaging over the three orthogonal main crystallographic directions. The SRO parameters
of all superstructures are summarized in Table \ref{table_SROs}.
\begin{table*}
\caption{The Warren-Cowley pair short range order parameters (SROs) up to the 7th neighboring shell
for each SQS supercell considered in this work.}
\label{table_SROs}
\begin{ruledtabular}
\begin{tabular}{ldddddddd}
str.$\setminus$shell & \multicolumn{1}{c}{\text{number of atoms}} & 1 & 2 & 3 & 4 & 5 & 6 & 7 \\
\hline
L1$_0$& 8 &-1.0&-1.0&-1.0&-1.0&-1.0&-1.0&-1.0\\
(2$\times$2$\times$2)& 16 &-0.16&0.0&-0.16&1.0&-0.16&0.0&-0.16\\
(2$\times$3$\times$2)& 24 &-0.11&0.0&-0.08&0.33&-0.06&-0.08&0.03\\
(4$\times$3$\times$2)& 48 &0.0&0.0&0.0&0.0&-0.06&0.0&0.0\\
(4$\times$3$\times$2)$^{\ast}$\footnotemark[1]& 48 &-0.14&0.28&-0.10&0.14&0.0&-0.17&-0.01\\
C1-(2$\times$2$\times$2)& 64 &-0.33&1.0&-0.33&1.0&-0.33&1.0&-0.33\\
C3-(2$\times$2$\times$2)& 64 &0.0&-1.0&0.0&-1.0&0.0&-1.0&0.0\\
B1-$(2\times 2\times2)$\footnotemark[2]& 64 &0.0&0.0&0.0&-0.33&0.0&0.0&0.0\\
(4$\times$3$\times$4)& 96 &0.0&0.0&0.0&0.0&0.0&0.0&0.0\\
(4$\times$4$\times$3)& 96 &0.0&0.0&-0.01&-0.01&-0.01&0.0&0.0\\
(4$\times$4$\times$4)& 128 &0.0&0.0&0.0&0.0&0.0&0.0&0.0\\
\end{tabular}
\end{ruledtabular}
\footnotetext[1]{The $\ast$ marks a different atomic configuration in the supercell.}
\footnotetext[2]{The supercell was obtained by von Pezold {\it et} al. in Ref.\cite{VonPezold2010}.}
\end{table*}
In the case of the $(2\times2\times2)$, $(2\times3\times2)$ and $(4\times4\times4)$ supercells,
those atomic configrations have been chosen that resulted closest to randomness in our approximation,
i.e. almost vanishing SRO parameters up to the seventh neighbor shell. The larger SROs in case of the
$(2\times2\times2)$ supercell are the consequence of the low configurational freedom in the supercell
and indicate less perfection in the randomness. In the case of the $(4\times3\times2)$ supercell size
two different atomic configurations have been considered with very different SRO parameters.
The $\ast$ marks the SQS structure that is less random. The calculated SROs of the C1-$(2\times2\times2)$
and C3-$(2\times2\times2)$ structures show alternating systematics that is related to the used
construction strategy. The SRO parameters deviate considerably from zero in these two cases. For example,
the cubic symmetric C3-$(2\times2\times2)$ shows perfect ordering in every second neighboring shell.
In comparing the SRO values in Table \ref{table_SROs}, the $(4\times4\times4)$ supercell gives
unambiguously the closest model of a totally random(pseudo-)binary alloy in our SQS approximation.

The structural optimization of these supercells resulted in slight structural distorsions, what are
summarized in Table \ref{table_structure}.
\begin{table}
\caption{The optimized structural parameters of Ti$_{0.5}$Al$_{0.5}$N obtained with different supercells from Table \ref{table_SROs}.}
\label{table_structure}
\begin{ruledtabular}
\begin{tabular}{lcccccc}
str. & $a(\text{\AA})$ & $b(\text{\AA})$ &  $c(\text{\AA})$ & $\measuredangle(b,c)$ & $\measuredangle(a,c)$ & $\measuredangle(a,b)$ \\
\hline
L1$_0$& 4.17 & 4.17 & 4.24 & 90.00 & 90.00 & 90.00 \\
(2$\times$2$\times$2)& 4.18 & 4.18 & 4.22 & 59.72 & 59.72 & 59.73\\
(2$\times$3$\times$2)& 4.20 & 4.19 & 4.17 & 60.17 & 60.12 & 59.65 \\
(4$\times$3$\times$2)& 4.18 & 4.18 & 4.20 & 59.67 & 59.79 & 59.99\\
(4$\times$3$\times$2)$^{\ast}$\footnotemark[1]& 4.19 & 4.19 & 4.17 & 60.14 & 60.17 & 59.66\\
C1-(2$\times$2$\times$2)& 4.25 & 4.16 & 4.16 & 90.00 & 90.00 & 90.00 \\
C3-(2$\times$2$\times$2)& 4.18 & 4.18 & 4.18 & 90.00 & 90.00 & 90.00\\
B1-(2$\times$2$\times$2)\footnotemark[2]& 4.18 & 4.18 & 4.18 & 89.78 & 89.78 & 90.00\\
(4$\times$3$\times$4)& 4.20 & 4.15 & 4.18 & 60.26 & 59.91 & 60.38\\
(4$\times$4$\times$3)& 4.19 & 4.18 & 4.16 & 60.15 & 60.13 & 59.94\\
(4$\times$4$\times$4)& 4.18 & 4.18 & 4.19 & 59.93 & 60.08 & 60.00 \\
\end{tabular}
\end{ruledtabular}
\footnotetext[1]{The $\ast$ marks a different atomic configuration in the supercell.}
\footnotetext[2]{The supercell was obtained by von Pezold {\it et} al. in Ref.\cite{VonPezold2010}.}
\end{table}
Table \ref{table_structure} gives the size resolved lattice parameters of B1 Ti$_{0.5}$A$_{0.5}$lN
within the different SQS supercell models. The lattice parameters, especially the length of the cell
edges, show some noticeable deviation from the cubic structure, but only for the L1$_0$,
$(2\times2\times2)$ and C1-$(2\times2\times2)$ supercells. What correlates with the systematic alternation
of the large SRO parameters of these cells. This suggests that the observed structural deviation is related
to the low  degree of freedom of internal atomic arrangement. In the other supercells with higher
substitutional atomic disorder/randomness the cubic imperfection is nearly negligible.

\begin{figure}[h!]
\includegraphics[width=8.5cm]{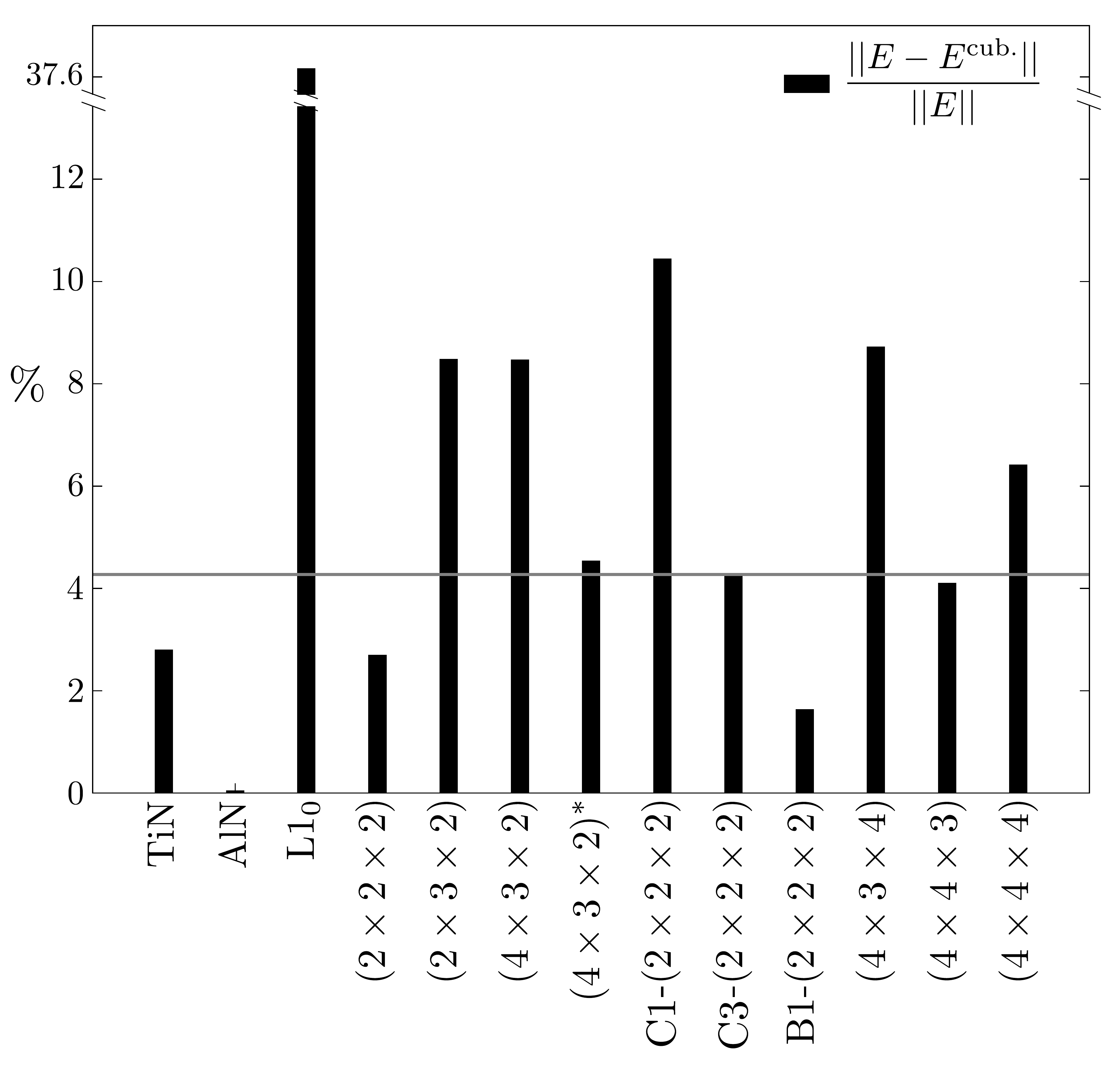}
\caption{\label{fig_01}(Color online)
Calculated euclidian norm deviations $||E-E^{\text{cub}}||/||E||$ obtained in the 21 dimensional space, see Eq.(\ref{elastic_vector}).}
\end{figure}

For each of these structures the full elastic tensor has been calculated. The elastic constants were
obtained independently, as 21 different distorsions were applied. The obtained elastic tensors are
summarized for all the structures in Appendix A. All the obtained tensors exhibit deviations from
a strict cubic symmetry, in which the principal non-vanishing elements should show the following
relationships,
\begin{eqnarray}
&\bar{C}_{11}=C_{11}=C_{22}=C_{33},\quad
\bar{C}_{12}=C_{12}=C_{13}=C_{23},\nonumber\\
&\bar{C}_{44}=C_{44}=C_{55}=C_{66}.
\label{cubic_sym}
\end{eqnarray}
Appendix B\ref{appendix_elastic_tensors} also lists the elastic tensors of bulk B1 TiN and AlN obtained
with the supercell size $(4\times4\times3)$, where one finds the cubic symmetry of elasticit constants with
numerical error. The values show good agreement with the literature data \cite{Chen2003,Mayrhofer2006}
obtained with different techniques. As the C3-$(2\times2\times2)$ supercell preserves the cubic symmetry,
its elastic tensor shows the cubic relationships in Eq.(\ref{cubic_sym}). The non-vanishing other elements
define the numerical accuracy, i.e. the average error $(6/463+6/182+6/156)/3$ should be around 3\%.
Since one gets the same 3\% numerical error in the case of bulk B1 TiN and a negligible one for B1 AlN,
we can assume that 3\% is numerical error threshold for all of our results through the following analysis.
One can read from the data in Appendix B that some of the SQS supercells result in large non-cubic elements
and large deviations between the principal cubic elastic constants, which means a breakdown of the cubic
symmetry relations in Eq.(\ref{cubic_sym}). Nevertheless, by the previously introduced projection we can
extract the closest cubic elastic tensors and calculate the distance variations
$||E-E^{\text{cub.}}||/||E||$. These deviations are shown in Fig.\ref{fig_01}. The required 9 elastic
constants are summarized in Table \ref{table_Cxy}, while the obtained projected cubic elastic constants
are listed in Table \ref{table_cubic_Cxy}.
\begin{table}[h!]
\caption{The calculated GGA elastic tensor elements of Ti$_{0.5}$Al$_{0.5}$N from Appendix B, whichs are
involved in the cubic projection, see Eqs.(\ref{cubic_proj}) and (\ref{cubic_average}).}
\label{table_Cxy}
\begin{ruledtabular}
\begin{tabular}{lccccccccc}
str.$\setminus$const.& $C_{11}$ & $C_{22}$ & $C_{33}$ & $C_{12}$ & $C_{13}$ &  $C_{23}$ & $C_{44}$ & $C_{55}$ & $C_{66}$ \\
\hline
TiN   & 617 & 618 & 618 & 123 & 123 & 123 & 178 & 178 & 178\\
AlN   & 402 & 402 & 402 & 157 & 157 & 157 & 300 & 300 & 300\\
L1$_0$& 409 & 409 & 332 & 183 & 197 & 197 & 100 & 100 & 120\\
(2$\times$2$\times$2)& 469 & 488 & 469 & 148 & 151 & 148 & 210 & 208 & 210\\
(2$\times$3$\times$2)& 429 & 388 & 443 & 173 & 164 & 169 & 187 & 203 & 188\\
(4$\times$3$\times$2)& 436 & 453 & 428 & 161 & 160 & 160 & 188 & 186 & 189\\
(4$\times$3$\times$2)$^{\ast}$\footnotemark[1]& 477 & 445 & 474 & 144 & 155 & 149 & 210 & 215 & 199\\
C1-(2$\times$2$\times$2)& 385 & 495 & 495 & 164 & 164 & 136 & 222 & 183 & 183\\
C3-(2$\times$2$\times$2)& 462 & 462 & 462 & 156 & 156 & 156 & 182 & 182 & 182\\
B1-(2$\times$2$\times$2)\footnotemark[2]& 481 & 482 & 473 & 139 & 147 & 147 & 214 & 214 & 218\\
(4$\times$3$\times$4)& 431 & 478 & 472 & 148 & 153 & 148 & 216 & 196 & 194\\
(4$\times$4$\times$3)& 456 & 425 & 460 & 161 & 152 & 160 & 201 & 211 & 198\\
(4$\times$4$\times$4)& 457 & 462 & 444 & 149 & 156 & 156 & 202 & 203 & 200\\
\end{tabular}
\end{ruledtabular}
\footnotetext[1]{The $\ast$ marks a different atomic configuration in the supercell.}
\footnotetext[2]{The supercell was obtained by von Pezold {\it et} al. in Ref.\cite{VonPezold2010}.}
\end{table}
\begin{table}[h!]
\caption{The projected principal cubic elastic constants and
the derived Zener's elastic anisotropy $\bar{A}=2\bar{C}_{44}/(\bar{C}_{11}-\bar{C}_{12})$
of Ti$_{0.5}$Al$_{0.5}$N obtained with the different structural models.}
\label{table_cubic_Cxy}
\begin{ruledtabular}
\begin{tabular}{lcccc}
str.$\setminus$const.& $\bar{C}_{11}$ & $\bar{C}_{12}$ & $\bar{C}_{44}$ & $\bar{A}$ \\
\hline
L1$_0$ & 384 & 193 & 107 & 1.12\\
(2$\times$2$\times$2)& 475 & 149 & 209 & 1.28\\
(2$\times$3$\times$2)& 420 & 169 & 193 & 1.53\\
(4$\times$3$\times$2)& 439 & 160 & 188 & 1.35\\
(4$\times$3$\times$2)$^{\ast}$\footnotemark[1]& 465 & 149 & 208 & 1.32\\
C1-(2$\times$2$\times$2)& 459 & 155 & 196 & 1.29\\
C3-(2$\times$2$\times$2)& 462 & 156 & 182 & 1.19\\
B1-(2$\times$2$\times$2)\footnotemark[2]& 479 & 144 & 215 & 1.29\\
(4$\times$3$\times$4)& 460 & 150 & 202 & 1.30\\
\Xhline{2\arrayrulewidth}
\bf{(4$\times$4$\times$3)}&\bf{447} & \bf{158} & \bf{203} & \bf{1.40}\\
\Xhline{2\arrayrulewidth}
(4$\times$4$\times$4)& 454 & 154 & 202 & 1.34\\
\end{tabular}
\end{ruledtabular}
\footnotetext[1]{The $\ast$ marks a different atomic configuration in the supercell.}
\footnotetext[2]{The supercell was obtained by von Pezold {\it et} al. in Ref.\cite{VonPezold2010}.}
\end{table}
Since the C3-$(2\times2\times2)$ supercell should have cubic symmetry, its
$||E-E^{\text{cub.}}||/||E||$ value defines the numerical threshold for the deviations, which
is around 4.3\%. Thus, only the $(2\times2\times2)$, C3-$(2\times2\times2)$, B1-$(2\times2\times2)$
and $(4\times4\times3)$ supercells give cubic symmetry within the most general 21 dimensional vector
space related to the 21 elastic constants. The $(4\times3\times2)^{\ast}$ and $(4\times4\times4)$
are the candidates to exhibit closely to cubic symmetry from elastic point of view. The ordered
L1$_0$ structure results in the largest deviation from cubic symmetry. While the $(2\times2\times2)$
supercell with relative large SRO parameters fulfills the cubic requirement, the larger and perfectly
random $(4\times3\times4)$ supercell does not. In general it underlines the importance of applying
supercells designed with the inclusion of symmetry in modeling anisotropic tensorial properties of
alloys. Namely, the SRO parameters or the atomic configuration should be optimized in such a way
as to support also the point group symmetry. From Fig. \ref{fig_01} with including the SRO parameters
we conclude that among the tested supercell structures our $(4\times4\times3)$ model should be taken
as the closest SQS model to study the elasticity in cubic Ti$_{0.5}$Al$_{0.5}$N. Thus, we conclude
that the accurate elastic constants of Ti$_{0.5}$Al$_{0.5}$N alloy are $C_{11}=447$ GPa, $C_{12}=158$
GPa and $C_{44}=203$ GPa within 3\% of numerical error. We also see, that with using very {\it ad-hoc}
or inadequate structures, such as the L1$0$, in predicting elastic constants of Ti$_{0.5}$Al$_{0.5}$N
one faces with large 22-50\% errors.

\begin{figure}[h!]
\includegraphics[width=8.5cm]{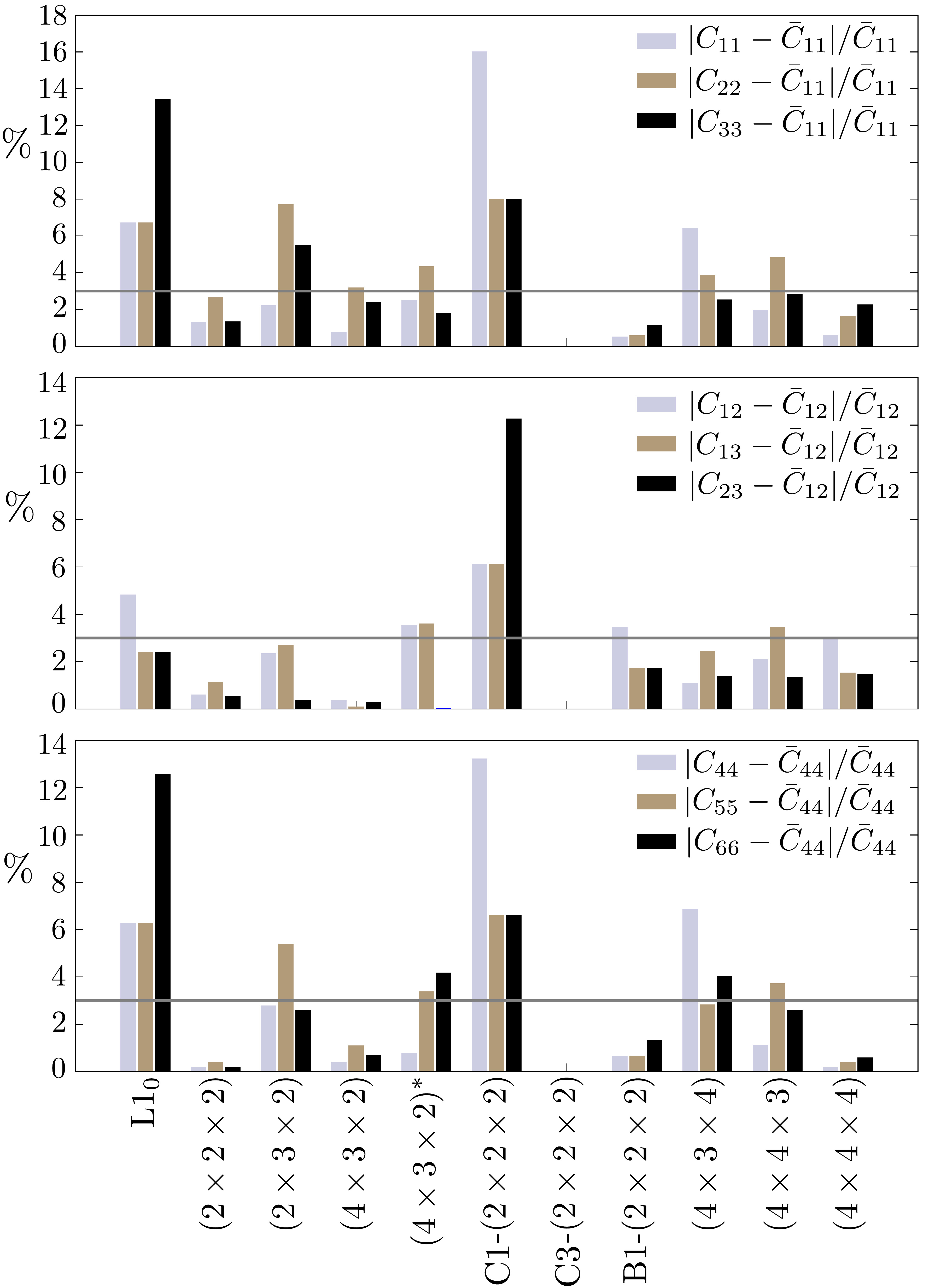}
\caption{\label{fig_02}(Color online)
Comparison of the calculated elastic tensor elements with the projected principal cubic elastic
constants in Ti$_{0.5}$Al$_{0.5}$N.}
\end{figure}

The projection technique allows us to evaluate the supercells in a smaller, 9 dimensional vector space.
In the following we consider only the nine elastic constants of $C_{11}$, $C_{22}$, $C_{33}$, $C_{12}$,
$C_{13}$, $C_{23}$, $C_{44}$, $C_{55}$ and $C_{66}$. These elastic constants are given in
Table \ref{table_Cxy}. The deviations of these constants from the projected cubic elastic constants
are shown in Fig. \ref{fig_02}.
In this figure the three columns for each supercell give the deviations along the three orthogonal
directions, [100], [010] and [001]. One can see in the figure, where the horizontal lines show our
3\% error threshold, that in the 9 dimensional space only three supercells, the $(2\times2\times2)$,
C3-$(2\times2\times)$ and $(4\times4\times4)$ give cubic elastic symmetry. Similarly to Fig. \ref{fig_01}
the $(4\times4\times3)$ supercells performs very well, while the totally random $(4\times3\times4)$
does not. Accordingly, Fig. \ref{fig_02} correlates quite well with Fig. \ref{fig_01}, namely we see
the same set of structures that performing perfectly good or bad. This leads us to the conclusion that
one can analyze the performance of the supercells in describing cubic elasticity within
this 9 dimensional subspace, too. This means a great reduction in the computational cost, since only
9 elements have to be calculated to measure the representation of elasticity.
By the way, the analysis in this 9 dimensional subspace might result in another best approximate
superstructure, like in this study. Fig. \ref{fig_02} shows clearly, that the $(4\times4\times4)$
supercell results in a somewhat better representation of cubic elastic symmetry in this space.

\begin{figure}[h!]
\includegraphics[width=8.5cm]{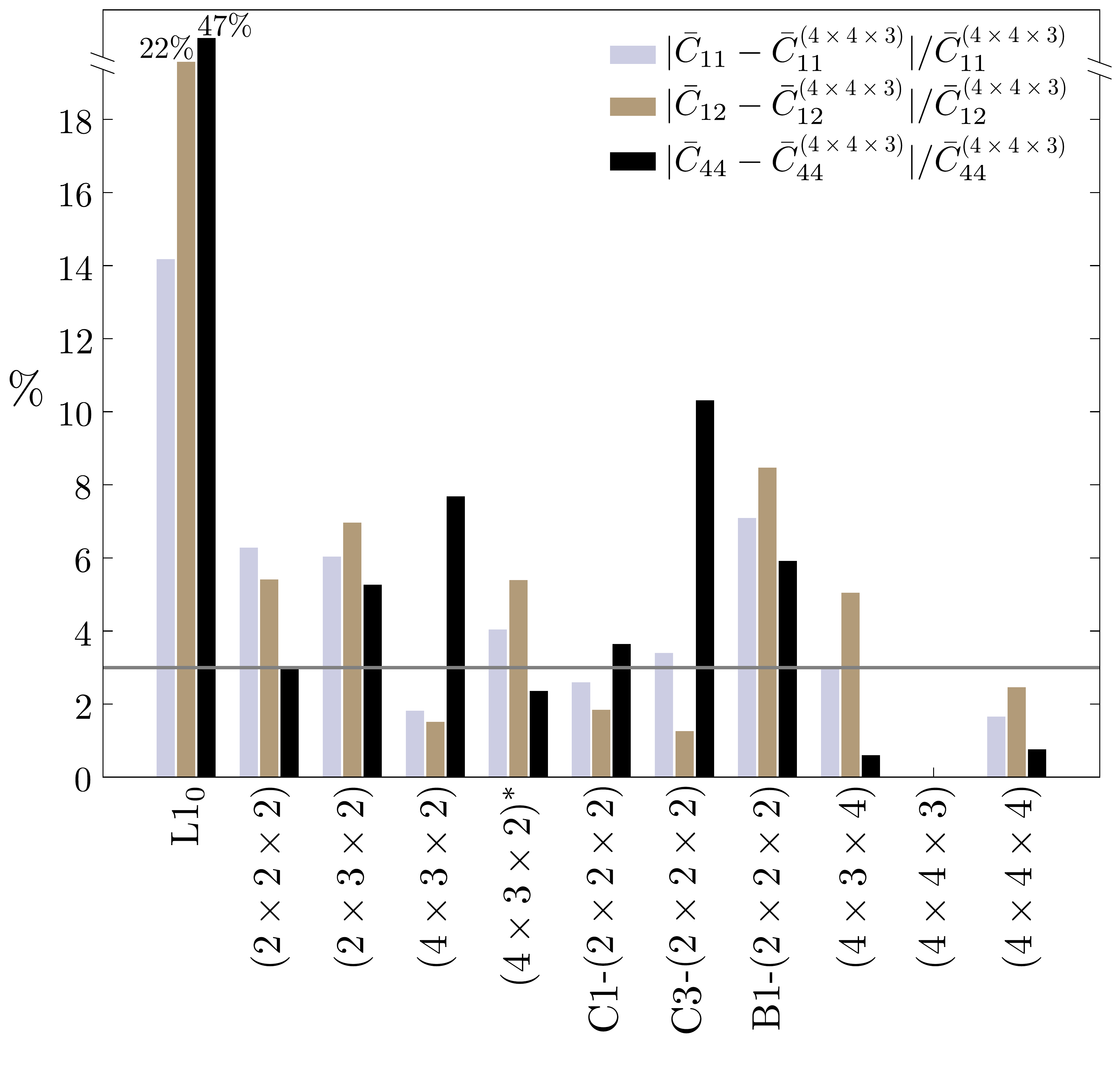}
\caption{\label{fig_03}(Color online)
The calculated projected cubic elastic constants of Ti$_{0.5}$Al$_{0.5}$N relative to the values obtained
for the $(4\times4\times3)$ SQS model.}
\end{figure}

However, this small discrepancy between the two previously performed analysis, within the full 21 and 9
dimensional spaces can be resolved by comparing the derived projected cubic elastic constants of the
supercells. This comparison is shown in Fig. \ref{fig_03}.
In Fig. \ref{fig_03} the relative deviations of elastic constants are plotted with respect to the
values obtained for the $(4\times4\times3)$ SQS in correspondence with the conclusion from
Fig.\ref{fig_02}. As one can see, the $(4\times3\times4)$ and $(4\times4\times4)$ supercells actually
result in the same cubic elastic constants within the 3\% numerical error. An interesting fact is that the
values in Fig. \ref{fig_03} should correlate with the corresponding relative differences in Fig. \ref{fig_01}.
See, for example, the big difference between the cases of $(4\times4\times4)$ and B1-$(2\times2\times2)$.
Accordingly, Fig. \ref{fig_03} concludes the convergency of the cubic elastic constants of Ti$_{0.5}$Al$_{0.5}$N
with respect to differently shaped and sized supercell models. Accordingly, the projected cubic elastic constants
can be used to predict elasticity of cubic alloys.

\begin{figure}[h!]
\includegraphics[width=8.5cm]{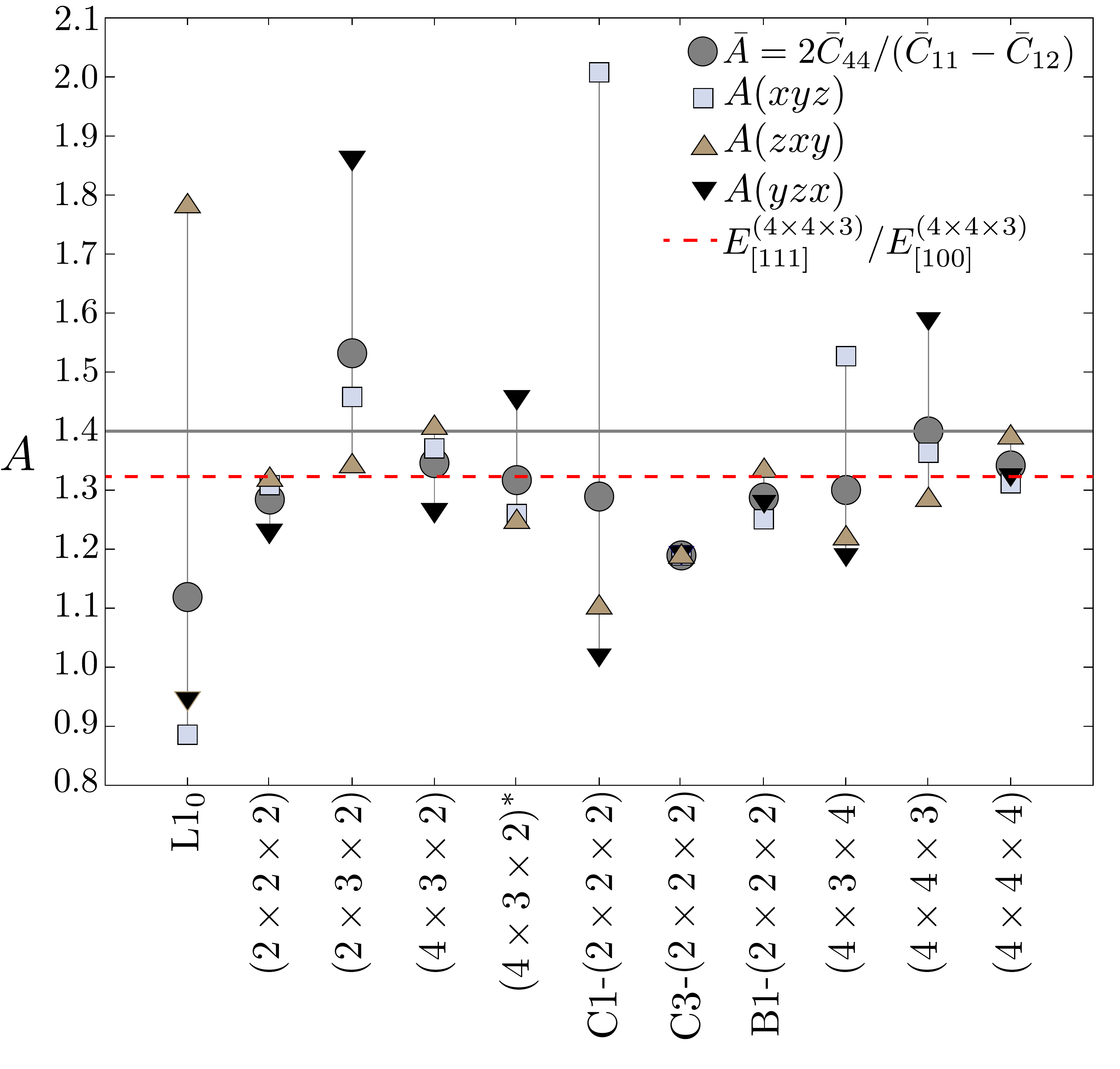}
\caption{\label{fig_04}(Color online)
Zener's elastic anisotropy values in Ti$_{0.5}$Al$_{0.5}$N for each structural models considered in this
study. The horizontal solid line shows the value of $\bar{A}$ obtained for the $(4\times4\times3)$ SQS.}
\end{figure}

Since the elastic anisotropy in TiAlN alloys has a huge impact on the materials mechanical properties
\cite{Tasnadi2010}, an accurate prediction of the Zener's elastic anisotropy is of a big importance. Using
the projected cubic elastic constants one can derive the Zener's elastic anisotropy via
\begin{equation}
\bar{A}=\frac{2\bar{C}_{44}}{\bar{C}_{11}-\bar{C}_{12}}.
\end{equation}
The derived values are listed in Table \ref{table_cubic_Cxy} and plotted in Fig. \ref{fig_04}.
The 3\% numerical error accumulates in the nominator and results in the approximate 5\% difference between
the two elastic anisotropy values obtained with the best $(4\times4\times4)$ and $(4\times4\times3)$ supercells.
Thus, the Zener's elastic anisotropy in Ti$_{0.5}$Al$_{0.5}$N should have the value of A=1.40 with around 6\%
numerical error. Fig.\ref{fig_04} shows not only the cubic projected elastic anisotropy values but also their
variation along the three orthogonal directions, [100], [010] and [001], using the data from Table \ref{table_Cxy}.
For example, in the [100] direction one has $A(x,y,z)=2C_{44}/(C_{11}-C{12})$ while in the [010],
$A(yzx)=2C_{66}/(C_{33}-C_{13})$. These orientational variations should vanish in case of true cubic
symmetry. However, as the figure shows one may get a large orientation dependence ($\approx 55\%$,
see C1-$(2\times2\times2)$) for a supercell being far from fulfilling cubic point symmetry. The sizes of
the variations should correlate with the deviations shown in Fig. \ref{fig_02}. Accordingly,
Fig. \ref{fig_04} gives a similar way to analyze the performance of the supercells in modeling elasticity
of cubic systems.

In the Reuss averaging method, with assumed uniform stress distribution, the strain ratio
$\epsilon_{[200]}/\epsilon_{[111]}=E_{[111]}/E_{[100]}$, where $E_{[hkl]}$ denotes the directional
Young's elastic moduli, can be applied to estimate elastic anisotropy experimentally in cubic materials.
Using our most accurate supercell model of $(4\times4\times3)$ the strain ratio $\epsilon_{[200]}/\epsilon_{[111]}$
has the value of 1.32 in Ti$_{0.5}$Al$_{0.5}$N. It is also shown in Fig.\ref{fig_04}. This value deviates
from our $\bar{A}=1.40$ value less than the 6\% numerical error. The elasticity of polycrystalline
Ti$_{0.5}$Al$_{0.5}$N can be discussed in terms of the Reuss and Voigt bulk ($B_{\text{R}}, B_{\text{V}}$)
and shear moduli ($G_{\text{R}}, G_{\text{V}}$) and also the derived Young's modulus ($E_{\text{V/G}}$) and
Poisson ratio ($\nu_{\text{V/G}}$),
\begin{widetext}
\begin{eqnarray}
B_{\text{V}}=\frac{(C_{11}+C_{22}+C_{33})+2(C_{12}+C_{13}+C{23})}{9},\nonumber\\
G_{\text{V}}=\frac{(C_{11}+C_{22}+C_{33})-(C_{12}+C_{13}+C{23})+3(C_{44}+C_{55}+C_{66})}{15},\nonumber\\
B_{\text{R}}=\frac{1}{(S_{11}+S_{22}+S_{33})+2(S_{12}+S_{13}+S_{23})},\nonumber\\
G_{\text{R}}=\frac{15}{4(S_{11}+S_{22}+S_{33})-4(S_{12}+S_{13}+S_{23})+3(S_{44}+S_{55}+S_{66})},\nonumber\\
E=\frac{9BG}{3B+G},\quad
\nu=\frac{3B-2G}{6B+2G}
\end{eqnarray}
\end{widetext}
where $S_{ij}$ denotes the elastic compliances. These polycrystalline averaged quantities obtained for our
the supercell $(4\times4\times3)$, that approximates both the randomness and cubic symmetry accurately, are
summarized in Table \ref{polycrystalline_moduli}.
\begin{table}[h!]
\caption{The polycrystalline bulk (B), shear (G), Young (E) moduli in unit of GPa and the Poisson ratio
of Ti$_{0.5}$Al$_{0.5}$N obtained with the $(4\times4\times3)$ supercell.}
\label{polycrystalline_moduli}
\begin{ruledtabular}
\begin{tabular}{cccccccc}
B$_{\text{V}}$&B$_{\text{R}}$&G$_{\text{V}}$&G$_{\text{R}}$&E$_{\text{V}}$&E$_{\text{R}}$&$\nu_{\text{V}}$&$\nu_{\text{R}}$\\
\hline
254&254&180&174&437&425&0.21&0.22
\end{tabular}
\end{ruledtabular}
\end{table}
The values clearly show the cubic requirement of $B_{\text{R}}=B_{\text{V}}$.
%
\section{Summary}
In this study we discuss the performance of superstructures, including approximate special quasirandom
structure (SQS) supercells in modeling the elasticity of cubic B1 Ti$_{0.5}$Al$_{0.5}$N alloy.
Though the SQS approach provides a successful scheme to model and predict the thermodynamics
of alloys, the technique is not aiming to represent tensorial materials properties with
symmetry. Thus, its straightforward application can not provide an unambiguous description of elasticity in
random alloys.

Here, we applied a symmetry based projection technique to accurately predict the cubic elastic tensor of 
B1 Ti$_{0.5}$Al$_{0.5}$N alloy within the SQS approach. We derived from {\it ab-initio} calculations the
closest cubic elastic tensor of B1 Ti$_{0.5}$Al$_{0.5}$N by using several supercells. With the help of
these derived cubic projected elastic constants we presented a detailed analysis and comparison of
differently shaped and sized supercell models in describing elasticity of a system with cubic symmetry.
Thus, we accurately determined the cubic elastic constants of cubic Ti$_{0.5}$Al$_{0.5}$N. The $(4\times4\times3)$
supercell provided us the best model of both, randomness and elasticity, which resulted in
$C_{11}=447$ GPa, $C_{12}=158$ GPa and $C_{44}=203$ GPa for the cubic elastic constants with 3\% of
error and $A=1.40$ for Zener's elastic anisotropy with 6\% of error.

With the help of the obtained elastic tensors, each with 21 constants, our results established the
fact that supercells with good SRO parameters may include large non-cubic elastic constants and, on
the contrary, supercells with bad SRO parameters might approximate cubic elastic tensor fairly accurately.
We showed that using only 9 elements,
$C_{11}$, $C_{22}$, $C_{33}$, $C_{12}$, $C_{13}$, $C_{23}$, $C_{44}$, $C_{55}$ and $C_{66}$ constants
from the tensors, one can also adequately evaluate the supercell models and convergency of the results.
We also showed that, the deviations between the three equivalent Zener-type anisotropy factors,
oriented along the [100], [010] and [001] directions, confirm the same observation and establish a
measure of approximate cubic symmetry.

In summary, in this study we accurately predict cubic elastic constants of B1  Ti$_{0.5}$Al$_{0.5}$N alloy and
establish in general the importance of selecting proper SQS supercells with symmetry arguments to reliably
model elasticity of alloys. Furthermore, we suggest the calculation of nine elastic tensor elements -
$C_{11}$, $C_{22}$, $C_{33}$, $C_{12}$, $C_{13}$, $C_{23}$, $C_{44}$, $C_{55}$ and $C_{66}$, to evaluate
and analyse the performance of supercells in describing elasticity of alloys.

\hspace{1cm}
\\

%
\section{Acknowledgemnt}
This work was supported by the SSF project Designed Multicomponent coatings,
MultiFilms and the Swedish Research Council (VR). Calculations have been performed at Swedish
National Infrastructure for Computing (SNIC).
\appendix
\section{Appendix A: The atomic distributions and coordinates in the supercells}
The atomic distributions and coordinates relative to the supercells lattice parameters are listed in
Tables \ref{table_supercell_structures_1},\ref{table_supercell_structures_2}.
For the atomic distributions and coordinates in C1-$(2\times2\times2)$, C3-$(2\times2\times2)$ and
B1-$(2\times2\times2)$, see Ref.\cite{VonPezold2010}.
%
\begin{table*}
\caption{The internal atomic structures of the supercells in relative coordinates.}
\label{table_supercell_structures_1}
\begin{ruledtabular}
\begin{tabular}{rlrlrlrl}
\multicolumn{2}{c}{($2\times2\times2$)} &
\multicolumn{2}{c}{($2\times3\times2$)} &
\multicolumn{2}{c}{($4\times3\times2$)} &
\multicolumn{2}{c}{($4\times3\times2$)} \\
\hline
  Ti &  0   0   0   &Ti & 0   0   0   &Al & 0   0   0   &Al & 0   0   0   \\
  Al &  1/2 0   0   &Al & 1/2 0   0   &Al & 1/4 0   0   &Ti & 1/4 0   0   \\
  Ti &  0   1/2 0   &Ti & 0   1/3 0   &Al & 1/2 0   0   &Al & 1/2 0   0   \\
  Ti &  1/2 1/2 0   &Al & 0   2/3 0   &Ti & 3/4 0   0   &Ti & 3/4 0   0   \\
  Al &  0   0   1/2 &Ti & 1/2 1/3 0   &Al & 0   1/3 0   &Ti & 0   1/3 0   \\
  Ti &  1/2 0   1/2 &Ti & 1/2 2/3 0   &Ti & 0   2/3 0   &Al & 0   2/3 0   \\
  Al &  0   1/2 1/2 &Al & 0   0   1/2 &Ti & 1/4 1/3 0   &Al & 1/4 1/3 0   \\
  Al &  1/2 1/2 1/2 &Al & 1/2 0   1/2 &Al & 1/4 2/3 0   &Ti & 1/4 2/3 0   \\
     &              &Ti & 0   1/3 1/2 &Ti & 1/2 1/3 0   &Ti & 1/2 1/3 0   \\
     &              &Al & 0   2/3 1/2 &Ti & 1/2 2/3 0   &Ti & 1/2 2/3 0   \\
     &              &Al & 1/2 1/3 1/2 &Al & 3/4 1/3 0   &Ti & 3/4 1/3 0   \\
     &              &Ti & 1/2 2/3 1/2 &Ti & 3/4 2/3 0   &Al & 3/4 2/3 0   \\
     &              &   &             &Ti & 0   0   1/2 &Ti & 0   0   1/2 \\
     &              &   &             &Al & 1/4 0   1/2 &Ti & 1/4 0   1/2 \\
     &              &   &             &Ti & 1/2 0   1/2 &Al & 1/2 0   1/2 \\
     &              &   &             &Ti & 3/4 0   1/2 &Al & 3/4 0   1/2 \\
     &              &   &             &Ti & 0   1/3 1/2 &Al & 0   1/3 1/2 \\
     &              &   &             &Al & 0   2/3 1/2 &Ti & 0   2/3 1/2 \\
     &              &   &             &Al & 1/4 1/3 1/2 &Al & 1/4 1/3 1/2 \\
     &              &   &             &Al & 1/4 2/3 1/2 &Al & 1/4 2/3 1/2 \\
     &              &   &             &Al & 1/2 1/3 1/2 &Ti & 1/2 1/3 1/2 \\
     &              &   &             &Al & 1/2 2/3 1/2 &Ti & 1/2 2/3 1/2 \\
     &              &   &             &Ti & 3/4 1/3 1/2 &Al & 3/4 1/3 1/2 \\
     &              &   &             &Ti & 3/4 2/3 1/2 &Al & 3/4 2/3 1/2 \\

\end{tabular}
\end{ruledtabular}
\end{table*}
\begin{table*}
\caption{The internal atomic structures of the supercells in relative coordinates (continuation).}
\label{table_supercell_structures_2}
\begin{ruledtabular}
\begin{tabular}{rlrlrlrlrlrlrl}
\multicolumn{2}{c}{($4\times3\times4$)} &
\multicolumn{2}{c}{($4\times4\times3$)} &
\multicolumn{2}{c}{($4\times4\times4$)} \\
\hline
Ti & 0     0    0  &Al  &     0   0   0&Ti  &   0      0      0\\
Al & 1/4   0    0  &Al  &     1/4   0   0&Al  &  1/4     0      0\\
Ti & 1/2   0    0  &Al  &     3/4   0   0&Al  &  1/2     0      0\\
Ti & 3/4   0    0  &Al  &     0   0   1/3&Ti  &  3/4     0      0\\
Ti &  0   1/3   0  &Al  &     0   0   2/3&Ti  &   0     1/4     0\\
Ti &  0   2/3   0  &Al  &     1/4   0   1/3&Ti  &   0     1/2     0\\
Al & 1/4  1/3   0  &Al  &     1/4   0   2/3&Ti  &   0     3/4     0\\
Al & 1/4  2/3   0  &Al  &     1/2   0   1/3&Al  &  1/4    1/4     0\\
Ti & 1/2  1/3   0  &Al  &     1/2   0   2/3&Ti  &  1/4    1/2     0\\
Ti & 1/2  2/3   0  &Al  &     0   3/4   0&Ti  &  1/4    3/4     0\\
Ti & 3/4  1/3   0  &Al  &     1/4   1/4   0&Al  &  1/2    1/4     0\\
Ti & 3/4  2/3   0  &Al  &     1/2   3/4   0&Ti  &  1/2    1/2     0\\
Ti &  0    0   1/4 &Al  &     3/4   1/4   0&Ti  &  1/2    3/4     0\\
Al &  0    0   1/2 &Al  &     0   3/4   1/3&Ti  &  3/4    1/4     0\\
Al &  0    0   3/4 &Al  &     0   1/2   2/3&Ti  &  3/4    1/2     0\\
Al & 1/4   0   1/4 &Al  &     0   3/4   2/3&Al  &  3/4    3/4     0\\
Al & 1/4   0   1/2 &Al  &     1/4   1/4   1/3&Ti  &   0      0     1/4\\
Ti & 1/4   0   3/4 &Al  &     1/4   1/4   2/3&Al  &   0      0     1/2\\
Ti & 1/2   0   1/4 &Al  &     1/4   1/2   2/3&Al  &   0      0     3/4\\
Al & 1/2   0   1/2 &Al  &     1/2   1/4   1/3&Ti  &  1/4     0     1/4\\
Al & 1/2   0   3/4 &Al  &     1/2   3/4   1/3&Al  &  1/4     0     1/2\\
Al & 3/4   0   1/4 &Al  &     1/2   3/4   2/3&Ti  &  1/4     0     3/4\\
Ti & 3/4   0   1/2 &Al  &     3/4   1/4   1/3&Al  &  1/2     0     1/4\\
Ti & 3/4   0   3/4 &Al  &     3/4   1/4   2/3&Al  &  1/2     0     1/2\\
Ti &  0   1/3  1/4 &Ti  &     1/2   0.0   0&Ti  &  1/2     0     3/4\\
Al &  0   1/3  1/2 &Ti  &     3/4   0.0   1/3&Al  &  3/4     0     1/4\\
Al &  0   1/3  3/4 &Ti  &     3/4   0.0   2/3&Al  &  3/4     0    1/2\\
Ti &  0   2/3  1/4 &Ti  &     0   1/4   0&Al  &  3/4     0     3/4\\
Al &  0   2/3  1/2 &Ti  &     0   1/2   0&Ti  &   0     1/4    1/4\\
Al &  0   2/3  3/4 &Ti  &     1/4   1/2   0&Al  &   0     1/4    1/2\\
Al & 1/4  1/3  1/4 &Ti  &     1/4   3/4   0&Al  &   0     1/4    3/4\\
Al & 1/4  1/3  1/2 &Ti  &     1/2   1/4   0&Al  &   0     1/2    1/4\\
Ti & 1/4  1/3  3/4 &Ti  &     1/2   1/2   0& Ti  &   0     1/2    1/2\\
Al & 1/4  2/3  1/4 &Ti  &     3/4   1/2   0&Ti  &   0     3/4    1/4\\
Al & 1/4  2/3  1/2 &Ti  &     3/4   3/4   0&Ti  &   0     3/4    1/2\\
Ti & 1/4  2/3  3/4 &Ti  &     0   1/4   1/3&Al  &   0     3/4    3/4\\
Ti & 1/2  1/3  1/4 &Ti  &     0   1/2   1/3&Ti  &  1/4    1/4    1/4\\
Al & 1/2  1/3  1/2 &Ti  &     0   1/4   2/3&Ti  &  1/4    1/4    1/2\\
Al & 1/2  1/3  3/4 &Ti  &     1/4   1/2   1/3&Al  &  1/4    1/4    3/4\\
Ti & 1/2  2/3  1/4 &Ti  &     1/4   3/4   1/3&Ti  &  1/4    1/2    1/4\\
Al & 1/2  2/3  1/2 &Ti  &     1/4   3/4   2/3&Al  &  1/4    1/2    1/2\\
Al & 1/2  2/3  3/4 &Ti  &    1/2   1/2   1/3&Ti  &  1/4    1/2    3/4\\
Al & 3/4  1/3  1/4 &Ti  &     1/2   1/4   2/3&Ti  &  1/4    3/4    1/4\\
Ti & 3/4  1/3  1/2 &Ti  &     1/2   1/2   2/3&Al  &  1/4    3/4    1/2\\
Ti & 3/4  1/3  3/4 &Ti  &     3/4   1/2   1/3&Al  &  1/4    3/4    3/4\\
Al & 3/4  2/3  1/4 &Ti  &     3/4   3/4   1/3&Ti  &  1/2    1/4    1/4\\
Ti & 3/4  2/3  1/2 &Ti  &     3/4   1/2   2/3&Al  &  1/2    1/4    1/2\\
Ti & 3/4  2/3  3/4 &Ti  &     3/4   3/4   2/3&Ti  &  1/2    1/4    3/4\\
   &                    &&&Al  &  1/2    1/2    1/2\\
   &                    &&&Al  &  1/2    1/2    3/4\\
   &                    &&&Ti  &  1/2    3/4    1/4\\
   &                    &&&Al  &  1/2    3/4    1/2\\
   &                    &&&Al  &  1/2    3/4    3/4\\
   &                    &&&Ti  &  3/4    1/4    1/4\\
   &                    &&&Al  &  3/4    1/4    1/2\\
   &                    &&&Al  &  3/4    1/4    3/4\\
   &                    &&&Al  &  3/4    1/2    1/4\\
   &                    &&&Ti  &  3/4    1/2    1/2\\
   &                    &&&Al  &  3/4    1/2    3/4\\
   &                    &&&Ti  &  3/4    3/4    1/4\\
   &                    &&&Ti  &  3/4    3/4    1/2\\
   &                    &&&Al  &  3/4    3/4    3/4\\
\end{tabular}
\end{ruledtabular}
\end{table*}

\newpage
\section{Appendix B: Elastic tensors (in GPa) of cubic TiN, AlN and Ti$_{0.5}$Al$_{0.5}$N calculated for supercells from Table \ref{table_SROs}}
\label{appendix_elastic_tensors}
\subsubsection{Elastic tensor of B1 TiN}
\begin{equation}
\begin{pmatrix}
617 & 123 & 123 & -2 & -6 & -6 \\
    & 618 & 123 & -6 & -2 & -6 \\
    &     & 618 & -6 & -6 & -2 \\
    &     &     & 178& -4 & -4 \\
    &     &     &    &178 & -4 \\
    &     &     &    &    & 178
\end{pmatrix}
\nonumber
\end{equation}
\subsubsection{Elastic tensor of B1 AlN}
\begin{equation}
\begin{pmatrix}
402 & 157 & 157 &  0 &  0 &  0 \\
    & 402 & 157 &  0 &  0 &  0 \\
    &     & 402 &  0 &  0 &  0 \\
    &     &     &300 &  0 &  0 \\
    &     &     &    &300 &  0 \\
    &     &     &    &    & 300
\end{pmatrix}
\nonumber
\end{equation}
\subsubsection{Elastic tensor of L1$_0$ structure}
\begin{equation}
\begin{pmatrix}
409 & 183 & 197 & 44 & 44 & 49 \\
    & 409 & 197 & 44 & 44 & 49 \\
    &     & 332 & 45 & 45 & 49 \\
    &     &     & 100& 44 & 46 \\
    &     &     &    &100 & 46 \\
    &     &     &    &    & 120
\end{pmatrix}
\nonumber
\end{equation}
\subsubsection{Elastic tensor of $(2\times2\times2)$ SQS}
\begin{equation}
\begin{pmatrix}
469 & 148 & 151 &  -3  & -5  & -3 \\
    & 488 & 148 &  -3  &  3  & -3 \\
    &     & 469  & -3  &  -5 & -3 \\
    &     &      & 210 &  -4 & -4 \\
    &     &      &     & 208 & -4 \\
    &     &      &     &     &  210
\end{pmatrix}
\nonumber
\end{equation}
\subsubsection{Elastic tensor of $(2\times3\times2)$ SQS}
\begin{equation}
\begin{pmatrix}
429 & 173 & 164 &   2  &   4 &  6 \\
    & 388 & 169 &  15  &  16 &  4 \\
    &     & 443 &  11  &   9 &  16 \\
    &     &     & 187  &   9 &  8  \\
    &     &     &      & 203 &  9 \\
    &     &     &      &     &  188
\end{pmatrix}
\nonumber
\end{equation}
\subsubsection{Elastic tensor of $(4\times3\times2)$ SQS}
\begin{equation}
\begin{pmatrix}
436 & 161 & 160 &  12 &  11 &  25 \\
    & 453 & 160 &  4  &  15 &   1 \\
    &     & 428 &  13 &   3 &  8 \\
    &     &     & 188 &  12 &   9\\
    &     &     &     & 186 &   9\\
    &     &     &     &     & 189
\end{pmatrix}
\nonumber
\end{equation}
\subsubsection{Elastic tensor of $(4\times3\times2)^{\ast}$ SQS}
\begin{equation}
\begin{pmatrix}
477&144&155&-2&2&9\\
&445&149&-1&3&-14\\
&&474&3&-6&1\\
&&&210&0&2\\
&&&&215&1\\
&&&&&199
\end{pmatrix}
\nonumber
\end{equation}
\subsubsection{Elastic tensor of C1-$(2\times2\times2)$ structure}
\begin{equation}
\begin{pmatrix}
385 & 164 & 164 &  4 &  4 &  4 \\
    & 495 & 136 &  0 &  1 & 0 \\
    &     & 495 &  0 &  0 & 1 \\
    &     &     &  222 & 3 & 3 \\
    &     &     &       &183  & 2\\
    &     &     &       &     &  183
\end{pmatrix}
\nonumber
\end{equation}
\subsubsection{Elastic tensor of C3-$(2\times2\times2)$ structure}
\begin{equation}
\begin{pmatrix}
462 & 156 & 156 &   6 &   6 &   6 \\
    & 462 & 156 &   6 &   6 &   6 \\
    &     & 462 &   6 &   6 &   6 \\
    &     &     & 182 &   6 &   6 \\
    &     &     &     & 182 &   6 \\
    &     &     &     &     & 182
\end{pmatrix}
\nonumber
\end{equation}
\subsubsection{Elastic tensor of B1-$(2\times2\times2)$ structure from Ref.\cite{VonPezold2010}.}
\begin{equation}
\begin{pmatrix}
481 & 139 & 147 &  -1 &  -4 & -2 \\
    & 482 & 147 &  -4 &  -1 & -2 \\
    &     & 473 &  -1 &  -1 & -2 \\
    &     &     & 214 &  -1 & -1 \\
    &     &     &     & 214 & -1 \\
    &     &     &     &     & 201
\end{pmatrix}
\nonumber
\end{equation}
\subsubsection{Elastic tensor of $(4\times3\times4)$ SQS}
\begin{equation}
\begin{pmatrix}
431 & 148 & 153 &  -2 &  25 &  21 \\
    & 478 & 148 & -11 &  -10& -16 \\
    &     & 472 &   9 & -11 & 5  \\
    &     &     & 216 &   3 &  0 \\
    &     &     &     & 196  &-1 \\
    &     &     &     &      & 194
\end{pmatrix}
\nonumber
\end{equation}
\subsubsection{Elastic tensor of $(4\times4\times3)$ SQS}
\begin{equation}
\begin{pmatrix}
456 & 161.05& 152&1&2&4\\
    &425&160&7&4&1\\
    &   &460&3&1&9\\
    &   &   &201&3&5\\
    &   &   &   &211&3\\
    &   &   &    &  &198
\end{pmatrix}
\nonumber
\end{equation}
\subsubsection{Elastic tensor of $(4\times4\times4)$ SQS}
\begin{equation}
\begin{pmatrix}
457 & 149 & 156 &  -2 & 14 &  19\\
    & 462 & 156 & -11 &  -3 & -16 \\
    &      & 444 &  17 &  -5 &  1 \\
    &      &     &  202 &  0 &  -1 \\
    &      &     &      & 203 & -1 \\
    &      &     &       &     & 200

\end{pmatrix}
\nonumber
\end{equation}
%
%
\bibliographystyle{aps}

\begin{thebibliography}{26}
\expandafter\ifx\csname natexlab\endcsname\relax\def\natexlab#1{#1}\fi
\expandafter\ifx\csname bibnamefont\endcsname\relax
  \def\bibnamefont#1{#1}\fi
\expandafter\ifx\csname bibfnamefont\endcsname\relax
  \def\bibfnamefont#1{#1}\fi
\expandafter\ifx\csname citenamefont\endcsname\relax
  \def\citenamefont#1{#1}\fi
\expandafter\ifx\csname url\endcsname\relax
  \def\url#1{\texttt{#1}}\fi
\expandafter\ifx\csname urlprefix\endcsname\relax\def\urlprefix{URL }\fi
\providecommand{\bibinfo}[2]{#2}
\providecommand{\eprint}[2][]{\url{#2}}

\bibitem[{\citenamefont{H\"{O}RLING et~al.}()\citenamefont{H\"{o}rling,
  HULTMAN, ODEN, SJ\"{O}L\'EN, and KARLSSON}}]{Horling2005}
\bibinfo{author}{\bibfnamefont{A.}~\bibnamefont{H\"{o}rling}},
  \bibinfo{author}{\bibfnamefont{L.}~\bibnamefont{Hultman}},
  \bibinfo{author}{\bibfnamefont{M.}~\bibnamefont{Od\'en}},
  \bibinfo{author}{\bibfnamefont{J.}~\bibnamefont{Sj\"{o}l\'en}},
  \bibnamefont{and} \bibinfo{author}{\bibfnamefont{L.}~\bibnamefont{Karlsson}},
  \bibinfo{journal}{Surf. Coat. Technol.}
  \textbf{\bibinfo{volume}{191}},
  \bibinfo{pages}{384} (\bibinfo{year}{2002}).

\bibitem[{\citenamefont{Alling et~al.}(2007)\citenamefont{Alling, Ruban,
  Karimi, Peil, Simak, Hultman, and Abrikosov}}]{Alling2007}
\bibinfo{author}{\bibfnamefont{B.}~\bibnamefont{Alling}},
  \bibinfo{author}{\bibfnamefont{A.}~\bibnamefont{Ruban}},
  \bibinfo{author}{\bibfnamefont{A.}~\bibnamefont{Karimi}},
  \bibinfo{author}{\bibfnamefont{O.}~\bibnamefont{Peil}},
  \bibinfo{author}{\bibfnamefont{S.}~\bibnamefont{Simak}},
  \bibinfo{author}{\bibfnamefont{L.}~\bibnamefont{Hultman}}, \bibnamefont{and}
  \bibinfo{author}{\bibfnamefont{I.}~\bibnamefont{Abrikosov}},
  \bibinfo{journal}{Phys. Rev. B} \textbf{\bibinfo{volume}{75}}
  \bibinfo{pages}{045123} (\bibinfo{year}{2007}).

\bibitem[{\citenamefont{Mayrhofer et~al.}(2006)\citenamefont{Mayrhofer, Musics,
  and Schneider}}]{Mayrhofer_APL2006}
\bibinfo{author}{\bibfnamefont{P.}~\bibnamefont{Mayrhofer}},
  \bibinfo{author}{\bibfnamefont{D.}~\bibnamefont{Musics}}, \bibnamefont{and}
  \bibinfo{author}{\bibfnamefont{J.M.}~\bibnamefont{Schenider}},
  \bibinfo{journal}{Appl. Phys. Lett.} \textbf{\bibinfo{volume}{88}}
  \bibinfo{pages}{071922}, (\bibinfo{year}{2006}).

\bibitem[{\citenamefont{Alling et~al.}(2008)\citenamefont{Alling, Karimi,
  Hultman, and Abrikosov}}]{Alling2008}
\bibinfo{author}{\bibfnamefont{B.}~\bibnamefont{Alling}},
  \bibinfo{author}{\bibfnamefont{A.}~\bibnamefont{Karimi}},
  \bibinfo{author}{\bibfnamefont{L.}~\bibnamefont{Hultman}}, \bibnamefont{and}
  \bibinfo{author}{\bibfnamefont{I.~A.} \bibnamefont{Abrikosov}},
  \bibinfo{journal}{Appl. Phys. Lett.} \textbf{\bibinfo{volume}{92}},
  \bibinfo{pages}{071903} (\bibinfo{year}{2008}).

\bibitem[{\citenamefont{Alling et~al.}(2009)\citenamefont{Alling, Ode\'en,
  Hultman, and Abrikosov}}]{Alling2009}
\bibinfo{author}{\bibfnamefont{B.}~\bibnamefont{Alling}},
  \bibinfo{author}{\bibfnamefont{M.}~\bibnamefont{Od\'en}},
  \bibinfo{author}{\bibfnamefont{L.}~\bibnamefont{Hultman}}, \bibnamefont{and}
  \bibinfo{author}{\bibfnamefont{I.~A.} \bibnamefont{Abrikosov}},
  \bibinfo{journal}{Appl. Phys. Lett.} \textbf{\bibinfo{volume}{95}},
  \bibinfo{pages}{181906} (\bibinfo{year}{2009}).

\bibitem[{\citenamefont{Lind et~al.}(2003)\citenamefont{Lind, Fors\'en,
  Alling, Ghafoor, Tasn\'adi, Johnasson, Abrikosov and Od\'en}}]{Lind2011}
\bibinfo{author}{\bibfnamefont{H.} \bibnamefont{Lind}},
  \bibinfo{author}{\bibfnamefont{R.} \bibnamefont{Fors\'en}},
  \bibinfo{author}{\bibfnamefont{B.}~\bibnamefont{Alling}},
  \bibinfo{author}{\bibfnamefont{N.}~\bibnamefont{Ghafoor}},
  \bibinfo{author}{\bibfnamefont{F.}~\bibnamefont{Tasn\'adi}},
  \bibinfo{author}{\bibfnamefont{M.P.}~\bibnamefont{Johansson}},
  \bibinfo{author}{\bibfnamefont{I.A.}~\bibnamefont{Abrikosov}}, \bibnamefont{and}
  \bibinfo{author}{\bibfnamefont{M.}~\bibnamefont{Od\'en}},
  \bibinfo{journal}{Appl. Phys. Lett.}
  \textbf{\bibinfo{volume}{99}}, \bibinfo{pages}{091903} (\bibinfo{year}{2011}).

\bibitem[{\citenamefont{Tasn\'adi et~al.}(2010)\citenamefont{Tasn\'adi,
  Abrikosov, Rogström, Almer, Johansson, and Odén}}]{Tasnadi2010}
\bibinfo{author}{\bibfnamefont{F.}~\bibnamefont{Tasn\'adi}},
  \bibinfo{author}{\bibfnamefont{I.~A.} \bibnamefont{Abrikosov}},
  \bibinfo{author}{\bibfnamefont{L.}~\bibnamefont{Rogstr\"om}},
  \bibinfo{author}{\bibfnamefont{J.}~\bibnamefont{Almer}},
  \bibinfo{author}{\bibfnamefont{M.~P.} \bibnamefont{Johansson}},
  \bibnamefont{and} \bibinfo{author}{\bibfnamefont{M.}~\bibnamefont{Odén}},
  \bibinfo{journal}{Appl. Phys. Lett.} \textbf{\bibinfo{volume}{97}},
  \bibinfo{pages}{231902} (\bibinfo{year}{2010}).

\bibitem[{\citenamefont{Abrikosov et~al.}(2010)\citenamefont{Abrikosov,
  Knutsson, Alling, Tasn\'adi, Lind, Hultman, and Odén}}]{Abrikosov2011}
\bibinfo{author}{\bibfnamefont{I.A.}~\bibnamefont{Abrikosov}},
  \bibinfo{author}{\bibfnamefont{A.} \bibnamefont{Knutsson}},
  \bibinfo{author}{\bibfnamefont{B.}~\bibnamefont{Alling}},
  \bibinfo{author}{\bibfnamefont{F.}~\bibnamefont{Tasn\'adi}},
  \bibinfo{author}{\bibfnamefont{H.} \bibnamefont{Lind}},
  \bibinfo{author}{\bibfnamefont{L.} \bibnamefont{Hultman}},
  \bibnamefont{and} \bibinfo{author}{\bibfnamefont{M.}~\bibnamefont{Odén}},
  \bibinfo{journal}{Materials} \textbf{\bibinfo{volume}{4}},
  \bibinfo{pages}{1599} (\bibinfo{year}{2011}).

\bibitem[{\citenamefont{Ruban and Abrikosov}(2008)}]{Ruban2008}
\bibinfo{author}{\bibfnamefont{A.~V.} \bibnamefont{Ruban}} \bibnamefont{and}
  \bibinfo{author}{\bibfnamefont{I.~A.} \bibnamefont{Abrikosov}},
  \bibinfo{journal}{Rep. Prog. Phys.}
  \textbf{\bibinfo{volume}{71}}, \bibinfo{pages}{046501}
  (\bibinfo{year}{2008}).

\bibitem[{\citenamefont{Liu et~al.}(2005)\citenamefont{Liu, van~de Walle,
  Ghosh, and Asta}}]{Liu2005}
\bibinfo{author}{\bibfnamefont{J.}~\bibnamefont{Liu}},
  \bibinfo{author}{\bibfnamefont{A.}~\bibnamefont{van~de Walle}},
  \bibinfo{author}{\bibfnamefont{G.}~\bibnamefont{Ghosh}}, \bibnamefont{and}
  \bibinfo{author}{\bibfnamefont{M.}~\bibnamefont{Asta}},
  \bibinfo{journal}{Phys. Rev. B} \textbf{\bibinfo{volume}{72}}
  \bibinfo{pages}{144109} (\bibinfo{year}{2005}).

\bibitem[{\citenamefont{van~de Walle}(2008)}]{VandeWalle2008}
\bibinfo{author}{\bibfnamefont{A.}~\bibnamefont{van~de Walle}},
  \bibinfo{journal}{Nat. Mater.} \textbf{\bibinfo{volume}{7}},
  \bibinfo{pages}{455} (\bibinfo{year}{2008}).

\bibitem[{\citenamefont{Asta et~al.}(1991)\citenamefont{Asta, Wolverton,
  de~Fontaine, and Dreyss\'{e}}}]{Asta1991}
\bibinfo{author}{\bibfnamefont{M.}~\bibnamefont{Asta}},
  \bibinfo{author}{\bibfnamefont{C.}~\bibnamefont{Wolverton}},
  \bibinfo{author}{\bibfnamefont{D.}~\bibnamefont{de~Fontaine}},
  \bibnamefont{and}
  \bibinfo{author}{\bibfnamefont{H.}~\bibnamefont{Dreyss\'{e}}},
  \bibinfo{journal}{Phys. Rev. B} \textbf{\bibinfo{volume}{44}},
  \bibinfo{pages}{4907} (\bibinfo{year}{1991}).

\bibitem[{\citenamefont{Zunger et~al.}(1990)\citenamefont{Zunger, Wei,
  Ferreira, and Bernard}}]{Zunger1990}
\bibinfo{author}{\bibfnamefont{A.}~\bibnamefont{Zunger}},
  \bibinfo{author}{\bibfnamefont{S.-H.} \bibnamefont{Wei}},
  \bibinfo{author}{\bibfnamefont{L.}~\bibnamefont{Ferreira}}, \bibnamefont{and}
  \bibinfo{author}{\bibfnamefont{J.}~\bibnamefont{Bernard}},
  \bibinfo{journal}{Phys. Rev. Lett.} \textbf{\bibinfo{volume}{65}},
  \bibinfo{pages}{353} (\bibinfo{year}{1990}).

\bibitem[{\citenamefont{Tasn\'adi et~al.}(2010)\citenamefont{Tasn\'adi, Alling,
  H\"oglund, Wingqvist, Birch, Hultman, and
  Abrikosov}}]{Tasnadi2010PRL}
\bibinfo{author}{\bibfnamefont{F.}~\bibnamefont{Tasn\'adi}},
  \bibinfo{author}{\bibfnamefont{B.}~\bibnamefont{Alling}},
  \bibinfo{author}{\bibfnamefont{C.}~\bibnamefont{H\"oglund}},
  \bibinfo{author}{\bibfnamefont{G.}~\bibnamefont{Wingqvist}},
  \bibinfo{author}{\bibfnamefont{J.}~\bibnamefont{Birch}},
  \bibinfo{author}{\bibfnamefont{L.}~\bibnamefont{Hultman}}, \bibnamefont{and}
  \bibinfo{author}{\bibfnamefont{I.~A.} \bibnamefont{Abrikosov}},
  \bibinfo{journal}{Phys. Rev. Lett.} \textbf{\bibinfo{volume}{104}},
  \bibinfo{pages}{137601} (\bibinfo{year}{2010}).

\bibitem[{\citenamefont{Wingqvist et~al.}(2010)\citenamefont{Wingqvist,
  Tasn\'adi, Zukauskaite, Birch, Arwin, and Hultman}}]{Wingqvist2010}
\bibinfo{author}{\bibfnamefont{G.}~\bibnamefont{Wingqvist}},
  \bibinfo{author}{\bibfnamefont{F.}~\bibnamefont{Tasn\'adi}},
  \bibinfo{author}{\bibfnamefont{A.}~\bibnamefont{Zukauskaite}},
  \bibinfo{author}{\bibfnamefont{J.}~\bibnamefont{Birch}},
  \bibinfo{author}{\bibfnamefont{H.}~\bibnamefont{Arwin}}, \bibnamefont{and}
  \bibinfo{author}{\bibfnamefont{L.}~\bibnamefont{Hultman}},
  \bibinfo{journal}{Appl. Phys. Lett.} \textbf{\bibinfo{volume}{97}},
  \bibinfo{pages}{112902} (\bibinfo{year}{2010}).

\bibitem[{\citenamefont{Mayrhofer et~al.}(2006)\citenamefont{Mayrhofer, Music,
  and Schneider}}]{Mayrhofer2006}
\bibinfo{author}{\bibfnamefont{P.~H.} \bibnamefont{Mayrhofer}},
  \bibinfo{author}{\bibfnamefont{D.}~\bibnamefont{Music}}, \bibnamefont{and}
  \bibinfo{author}{\bibfnamefont{J.~M.} \bibnamefont{Schneider}},
  \bibinfo{journal}{J. Appl. Phys.} \textbf{\bibinfo{volume}{100}},
  \bibinfo{pages}{094906} (\bibinfo{year}{2006}).

\bibitem[{\citenamefont{Tasnádi et~al.}(2009)\citenamefont{Tasn\'adi,
  Abrikosov, and Katardjiev}}]{Tasnadi2009}
\bibinfo{author}{\bibfnamefont{F.}~\bibnamefont{Tasn\'adi}},
  \bibinfo{author}{\bibfnamefont{I.~A.} \bibnamefont{Abrikosov}},
  \bibnamefont{and}
  \bibinfo{author}{\bibfnamefont{I.}~\bibnamefont{Katardjiev}},
  \bibinfo{journal}{Appl. Phys. Lett.} \textbf{\bibinfo{volume}{94}},
  \bibinfo{pages}{151911} (\bibinfo{year}{2009}).

\bibitem[{\citenamefont{M\"{a}der and Zunger}(1995)}]{Mader1995}
\bibinfo{author}{\bibfnamefont{K.}~\bibnamefont{M\"{a}der}} \bibnamefont{and}
  \bibinfo{author}{\bibfnamefont{A.}~\bibnamefont{Zunger}},
  \bibinfo{journal}{Phys. Rev. B} \textbf{\bibinfo{volume}{51}},
  \bibinfo{pages}{10462} (\bibinfo{year}{1995}).

\bibitem[{\citenamefont{Bernardini et~al.}(2001)\citenamefont{Bernardini, and
  Bernardini}}]{Bernardini2001}
\bibinfo{author}{\bibfnamefont{F.}~\bibnamefont{Bernardini}},
  \bibnamefont{and}\bibinfo{author}{\bibfnamefont{V.}~\bibnamefont{Fiorentini}},
  \bibinfo{journal}{Phys. Rev. B} \textbf{\bibinfo{volume}{64}},
  \bibinfo{pages}{085207} (\bibinfo{year}{2001}).

\bibitem[{\citenamefont{von Pezold et~al.}(2010)\citenamefont{von Pezold, Dick,
  Fri\'{a}k, and Neugebauer}}]{VonPezold2010}
\bibinfo{author}{\bibfnamefont{J.}~\bibnamefont{von Pezold}},
  \bibinfo{author}{\bibfnamefont{A.}~\bibnamefont{Dick}},
  \bibinfo{author}{\bibfnamefont{M.}~\bibnamefont{Fri\'{a}k}},
  \bibnamefont{and}
  \bibinfo{author}{\bibfnamefont{J.}~\bibnamefont{Neugebauer}},
  \bibinfo{journal}{Phys. Rev. B} \textbf{\bibinfo{volume}{81}}
  (\bibinfo{year}{2010}).

\bibitem[{\citenamefont{Cowley}(1950)}]{Cowley1950}
\bibinfo{author}{\bibfnamefont{J.}~\bibnamefont{Cowley}},
  \bibinfo{journal}{Phys. Rev.} \textbf{\bibinfo{volume}{77}},
  \bibinfo{pages}{669} (\bibinfo{year}{1950}).

\bibitem[{\citenamefont{Metropolis et~al.}(1953)\citenamefont{Metropolis,
  Rosenbluth, Rosenbluth, Teller, and Teller}}]{Metropolis1953}
\bibinfo{author}{\bibfnamefont{N.}~\bibnamefont{Metropolis}},
  \bibinfo{author}{\bibfnamefont{A.~W.} \bibnamefont{Rosenbluth}},
  \bibinfo{author}{\bibfnamefont{M.~N.} \bibnamefont{Rosenbluth}},
  \bibinfo{author}{\bibfnamefont{A.~H.} \bibnamefont{Teller}},
  \bibnamefont{and} \bibinfo{author}{\bibfnamefont{E.}~\bibnamefont{Teller}},
  \bibinfo{journal}{J. Chem. Phys.}
  \textbf{\bibinfo{volume}{21}}, \bibinfo{pages}{1087} (\bibinfo{year}{1953}).

\bibitem[{\citenamefont{Vanderbilt}(1990)}]{Vanderbilt1990}
\bibinfo{author}{\bibfnamefont{D.}~\bibnamefont{Vanderbilt}},
  \bibinfo{journal}{Phys. Rev. B} \textbf{\bibinfo{volume}{41}},
  \bibinfo{pages}{7892} (\bibinfo{year}{1990}).

\bibitem[{\citenamefont{Giannozzi et~al.}(2009)\citenamefont{Giannozzi, Baroni,
  Bonini, Calandra, Car, Cavazzoni, Ceresoli, Chiarotti, Cococcioni, Dabo
  et~al.}}]{Giannozzi2009}
\bibinfo{author}{\bibfnamefont{P.}~\bibnamefont{Giannozzi}},
  \bibinfo{author}{\bibfnamefont{S.}~\bibnamefont{Baroni}},
  \bibinfo{author}{\bibfnamefont{N.}~\bibnamefont{Bonini}},
  \bibinfo{author}{\bibfnamefont{M.}~\bibnamefont{Calandra}},
  \bibinfo{author}{\bibfnamefont{R.}~\bibnamefont{Car}},
  \bibinfo{author}{\bibfnamefont{C.}~\bibnamefont{Cavazzoni}},
  \bibinfo{author}{\bibfnamefont{D.}~\bibnamefont{Ceresoli}},
  \bibinfo{author}{\bibfnamefont{G.~L.} \bibnamefont{Chiarotti}},
  \bibinfo{author}{\bibfnamefont{M.}~\bibnamefont{Cococcioni}},
  \bibinfo{author}{\bibfnamefont{I.}~\bibnamefont{Dabo}}, \bibnamefont{et~al.},
  \bibinfo{journal}{J. of Phys.: Condens. Matter}
  \textbf{\bibinfo{volume}{21}}, \bibinfo{pages}{395502}
  (\bibinfo{year}{2009}).

\bibitem[{\citenamefont{Perdew et~al.}(1996)\citenamefont{Perdew, Burke, and
  Ernzerhof}}]{Perdew1996}
\bibinfo{author}{\bibfnamefont{J.~P.} \bibnamefont{Perdew}},
  \bibinfo{author}{\bibfnamefont{K.}~\bibnamefont{Burke}}, \bibnamefont{and}
  \bibinfo{author}{\bibfnamefont{M.}~\bibnamefont{Ernzerhof}},
  \bibinfo{journal}{Phys. Rev. Lett.} \textbf{\bibinfo{volume}{77}},
  \bibinfo{pages}{3865} (\bibinfo{year}{1996}).

\bibitem[{\citenamefont{Monkhorst and Pack}(1976)}]{Monkhorst1976}
\bibinfo{author}{\bibfnamefont{H.}~\bibnamefont{Monkhorst}} \bibnamefont{and}
  \bibinfo{author}{\bibfnamefont{J.}~\bibnamefont{Pack}},
  \bibinfo{journal}{Phys. Rev. B} \textbf{\bibinfo{volume}{13}},
  \bibinfo{pages}{5188} (\bibinfo{year}{1976}).

\bibitem[{\citenamefont{Chen et~al.}(2003)\citenamefont{Chen, Zhao, Rodgers, and
  Tse}}]{Chen2003}
  \bibinfo{author}{\bibfnamefont{K.} \bibnamefont{Chen}},
  \bibinfo{author}{\bibfnamefont{L.R.} \bibnamefont{Zhao}},
  \bibinfo{author}{\bibfnamefont{J.}~\bibnamefont{Rodgers}}, \bibnamefont{and}
  \bibinfo{author}{\bibfnamefont{J.S.}~\bibnamefont{Tse}},
  \bibinfo{journal}{J. Phys. D: Appl. Phys} \textbf{\bibinfo{volume}{36}},
  \bibinfo{pages}{2725} (\bibinfo{year}{2003}).

\bibitem[{\citenamefont{Wentzcovitch et~al.}(1993)\citenamefont{Wentzcovitch,
  Martins, and Price}}]{Wentzcovitch1993}
\bibinfo{author}{\bibfnamefont{R.~M.} \bibnamefont{Wentzcovitch}},
  \bibinfo{author}{\bibfnamefont{J.~L.} \bibnamefont{Martins}},
  \bibnamefont{and} \bibinfo{author}{\bibfnamefont{G.~D.} \bibnamefont{Price}},
  \bibinfo{journal}{Phys. Rev. Lett.} \textbf{\bibinfo{volume}{70}},
  \bibinfo{pages}{3947} (\bibinfo{year}{1993}).

\bibitem[{\citenamefont{Nye}(1985)}]{Nye1985}
\bibinfo{author}{\bibfnamefont{J.~F.} \bibnamefont{Nye}},
  \emph{\bibinfo{title}{Physical Properties of Crystals: Their Representation
  by Tensors and Matrices}} (\bibinfo{publisher}{Oxford University Press, USA},
  \bibinfo{year}{1985}), ISBN \bibinfo{isbn}{0198511655}.

\bibitem[{\citenamefont{Vitos}(2010)}]{Vitos2010}
\bibinfo{author}{\bibfnamefont{L.}~\bibnamefont{Vitos}},
  \emph{\bibinfo{title}{{Computational Quantum Mechanics for Materials
  Engineers: The EMTO Method and Applications (Engineering Materials and
  Processes)}}} (\bibinfo{publisher}{Springer}, \bibinfo{year}{2010}), ISBN
  \bibinfo{isbn}{1849966850}.

\bibitem[{\citenamefont{Browaeys and Chevrot}(2004)}]{Browaeys2004}
\bibinfo{author}{\bibfnamefont{J.~T.} \bibnamefont{Browaeys}} \bibnamefont{and}
  \bibinfo{author}{\bibfnamefont{S.}~\bibnamefont{Chevrot}},
  \bibinfo{journal}{Geophys. J. Int.}
  \textbf{\bibinfo{volume}{159}}, \bibinfo{pages}{667} (\bibinfo{year}{2004}).

\bibitem[{\citenamefont{Moakher and Norris}(2006)}]{Moakher2006}
\bibinfo{author}{\bibfnamefont{M.}~\bibnamefont{Moakher}} \bibnamefont{and}
  \bibinfo{author}{\bibfnamefont{A.~N.} \bibnamefont{Norris}},
  \bibinfo{journal}{J. Elasticity} \textbf{\bibinfo{volume}{85}},
  \bibinfo{pages}{215} (\bibinfo{year}{2006}).

\end{thebibliography}

%
\end{document}